\documentclass[aps,reprint,showpacs,floatfix,citeautoscript,longbibliography,superscriptaddress]{revtex4-2}
\usepackage{times}
\usepackage{graphicx}
\usepackage{dcolumn}
\usepackage{bm}
\usepackage[colorlinks=true, linkcolor=blue, citecolor=blue, urlcolor=blue, linktoc=page, bookmarks=false]{hyperref} 

\usepackage[figure]{hypcap}

\begin{document}

\preprint{APS/123-QED}

\title{Anisotropic spin filtering by an altermagnetic barrier in magnetic tunnel junctions}

\author{Boyuan Chi}
\affiliation{Beijing National Laboratory for Condensed Matter Physics, Institute of Physics, University of Chinese Academy of Sciences, Chinese Academy of Sciences, Beijing 100190, China}

\author{Leina Jiang}
\affiliation{Beijing National Laboratory for Condensed Matter Physics, Institute of Physics, University of Chinese Academy of Sciences, Chinese Academy of Sciences, Beijing 100190, China}

\author{Yu Zhu}
\affiliation{Beijing National Laboratory for Condensed Matter Physics, Institute of Physics, University of Chinese Academy of Sciences, Chinese Academy of Sciences, Beijing 100190, China}

\author{Guoqiang Yu}
\affiliation{Beijing National Laboratory for Condensed Matter Physics, Institute of Physics, University of Chinese Academy of Sciences, Chinese Academy of Sciences, Beijing 100190, China}
\affiliation{Songshan Lake Materials Laboratory, Dongguan, Guangdong 523808, China}

\author{Caihua Wan}
\affiliation{Beijing National Laboratory for Condensed Matter Physics, Institute of Physics, University of Chinese Academy of Sciences, Chinese Academy of Sciences, Beijing 100190, China}

\author{Xiufeng Han}
\altaffiliation{xfhan@iphy.ac.cn}
\affiliation{Beijing National Laboratory for Condensed Matter Physics, Institute of Physics, University of Chinese Academy of Sciences, Chinese Academy of Sciences, Beijing 100190, China}
\affiliation{Songshan Lake Materials Laboratory, Dongguan, Guangdong 523808, China}
\affiliation{Center of Materials Science and Optoelectronics Engineering, University of Chinese Academy of Sciences, Beijing 100049, China}

\begin{abstract}
  The spin filtering effect, distinct decaying lengths experienced by oppositely spin-polarized electrons in a magnetic barrier, generally occurs in ferromagnetic (FM) insulators or semiconductors. With the rise of altermagnetic (ALM) materials which exhibit similar capability of spin-polarizing electrons with ferromagnets, it is a nature question whether the ALM insulators or semiconductors can also act as unique barriers for the spin splitting effect. Here, through first-principles calculations, we investigated the complex band structure of the ALM insulator FeF$_2$ and found that it possesses an anisotropic spin filtering effect: along the [001] direction of FeF$_2$, a current remains spin-neutral but has locally nonvanishing spin polarizations in the momentum space; moreover, along the [110] direction of FeF$_2$, a current will be globally spin-polarized by different attenuation lengths of oppositely spin-polarized electrons. Leveraging this anisotropic spin filtering effect, we designed two types of MTJs with the ALM barrier: ALM electrode/ALM insulator barrier/non-magnetic (NM) electrode and FM electrode/ALM insulator barrier/NM electrode, using RuO$_2$(001)/FeF$_2$/IrO$_2$ and CrO$_2$(110)/FeF$_2$/IrO$_2$ as the corresponding prototypes, respectively. We found that these two proposed MTJs exhibited the tunneling magnetoresistance (TMR) ratios of 216\% and 3956\%, by matching the conduction channels of the electrodes and the spin-resolved lowest decay rate of the barrier in the momentum space. Our work deepens and generalizes understanding toward the spin filtering effect for the rising ALM  insulators and semiconductors, and broadens applications of the AFM spintronics.
\end{abstract}

\maketitle

  The magnetic tunnel junction (MTJ), composed of two ferromagnetic (FM) electrodes and a sandwiched insulating barrier, is elementary for spintronic devices \cite{PLA-1975,PRL-74-1995,JMMM-1995}. The tunneling magnetoresistance (TMR) effect is the core of MTJ, making it suitable for magnetic random-access memories (MRAMs), magnetic sensors, and others spintronic devices \cite{S-2001,RMP-2004,IEEE-2006,JPCM-15-R1603-2003,JPCM-15-R109-2003}. To achieve reliable reading and writing, a higher TMR ratio is imperative. Initially, TMR was regarded dominantly determined by the spin polarization of FM electrodes. Therefore, various FM metals with high spin polarizations such as Heusler alloys have been paid plenty of attentions. Thereafter, the importance of symmetry filtering of crystallized barriers for tunneling was gradually realized \cite{PRL-85-2000}. By analyzing the evanescent states, MgO was predicted to exhibit weak attenuation effect for electrons with the $\Delta_1$ symmetry \cite{PRB-63-2001-1,PRB-63-2001-2}. Consequently, by combining Fe electrode with MgO, one successfully developed MTJs with large TMR ratios \cite{PRB-63-2001-1,PRB-63-2001-2,NM-3-2004-1,NM-3-2004-2}, manifesting the powerfulness of the barrier engineering for MTJs.

  Recently, the discovery of the antiferromagnetic (AFM) materials with spin splitting band structures, i.e., altermagnetic (ALM) materials \cite{PRX-12-031042-2022,PRX-12-040501-2022,N-2024}, has attracted researchers to explore their applications in magnetic devices, two-dimensional materials, superconductivity, etc. \cite{N-2024,npj2024,PRL-130-2023,PRB-109-2024-224430,PRB-107-2023,PRL-132-2024,PRB-110-2024-024503,arXiv10201,arXiv14620,NL-2024}. Focusing on the local spin polarization in momentum space, researchers have designed a novel all-antiferromagnetic tunnel junctions (AFMTJs) utilizing two ALM (or spin splitting AFM) electrodes instead of FM electrodes \cite{NC-2021,PRL-128-2022,N-613-485-2023,N-613-490-2023,PRB-108-2023,PRX-12-011028-2022,npj2024,arXiv02458}. Moreover, the TMR ratio can be further enhanced by matching the distribution of the lowest decay rates of the barrier with that of the conduction channels of the AFM electrodes in the reciprocal space \cite{arXiv03026}.

  In fact, the momentum-dependent spin splitting of ALM materials is not only in the $real$ part as reported in \cite{PRX-12-031042-2022,PRB-102-2020,PRM-2021,JPSJ-2019} but also in the $imaginary$ part of their band structures as shown below. Here, by analyzing the complex band structure of the ALM insulator FeF$_2$, we found a noticeable splitting in the decay rate $\kappa$ of the evanescent states for the two spin-channels. This feature is distinguished from that observed in conventional non-magnetic (NM) or trivial AFM insulators with the same decay rate for opposite spins. The splitting decay rates of evanescent states thereby result in the spin-resolved effective barrier for tunneling electrons with opposite spins, thus allowing ALM insulators to act as a spin filter with zero stray field. The generation of spin filtering effect without producing stray fields is a distinctive property of ALM materials, which makes them applicable in spin- and angle-resolved photoemission spectroscopy (SARPES) and enhancing the TMR effect, etc.
  
  Utilizing the spin filtering effect of ALM insulators, it is viable to design non-volatile MTJs with a magnetic compensated barrier. In this study, through first-principles calculations, we demonstrated the anisotropic spin filtering effect of ALM material FeF$_2$. Based on this effect, we designed two kinds of MTJ: ALM/ALM insulator barrier/NM and FM/ALM insulator barrier/NM MTJs. Specifically, we used RuO$_2$(001)/FeF$_2$/IrO$_2$ and CrO$_2$(110)/FeF$_2$/IrO$_2$ as the corresponding prototypes to study these two kinds of MTJs. The quantum transport calculations revealed that the two MTJs exhibit the TMR ratios of 216\% and 3956\%, respectively. The TMR effect in both arises from the concerted play of the spin filtering effect of FeF$_2$ and the spin-polarization of the FM and ALM electrodes, CrO$_2$ and RuO$_2$. The computational details are presented in the Supplementary Materials \cite{SM}. Due to the usage of NM electrode and ALM barrier with compensated magnetization, both MTJs do not require additional pinning structures, simplifying the device structures. Furthermore, the former MTJ benefits from zero stray field; the latter can be applied in magnetic sensor applications.

  In general, the transmission coefficient $T_\sigma(\boldsymbol{k}_\parallel)$ is usually factorized as \cite{JPCM-15-R1603-2003,PRB-69-2004-2,PRL-98-2007}:
  \begin{equation}
	\label{eq.1}
    T_\sigma(\boldsymbol{k}_\parallel)=t_L^\sigma(\boldsymbol{k}_\parallel)\exp[-2\kappa(\boldsymbol{k}_\parallel)d]t_R^\sigma(\boldsymbol{k}_\parallel)
  \end{equation}
  where $t_L^\sigma(\boldsymbol{k}_\parallel)$ and $t_R^\sigma(\boldsymbol{k}_\parallel)$ are the probabilities for an electron with spin $\sigma$ to transmit through the left and right electrode/barrier interfaces, respectively. $\kappa(\boldsymbol{k}_\parallel)$ is the lowest decay rate in the barrier and $d$ is the barrier width. 

  This equation applies to NM or trivial AFM barriers with spin degeneracy, such as MgO, NiO, and CoO, etc. However, for FM insulators, the lowest decay rate $\kappa$ is dependent on spin. Thus, Eq.(2) should be rewritten as:
  \begin{equation}
	\label{eq.2}
     T_\sigma(\boldsymbol{k}_\parallel)=t_L^\sigma(\boldsymbol{k}_\parallel)\exp[-2\kappa_\sigma(\boldsymbol{k}_\parallel)d]t_R^\sigma(\boldsymbol{k}_\parallel) 
  \end{equation}
  The spin-dependent $\kappa_\sigma(\boldsymbol{k}_\parallel)$ indicates that electrons with opposite spins experience different effective barrier heights during tunneling, and the barrier acts as a spin filter.

  An ALM material is a nontrivial AFM material that exhibits spin splitting in the absence of spin-orbit coupling. Therefore, for ALM insulators, like FeF$_2$, LaMnO$_3$, MnTe, et al., spin splitting occurs along certain high-symmetry directions. This indicates that the spin splitting can also manifest itself in the complex band along these directions as shown below, resulting in a spin-dependent $\kappa_\sigma(\boldsymbol{k}_\parallel)$, analogous to the case of FM insulators, as described by Eq. \hyperref[eq.2]{(2)}. As a result, when electrons tunnel along these spin-split directions of the ALM insulators, the current will be spin-polarized, and the ALM insulators can thus exhibit the spin filtering effect.

  \begin{figure}
	\centering
	\includegraphics[width=1.0\linewidth]{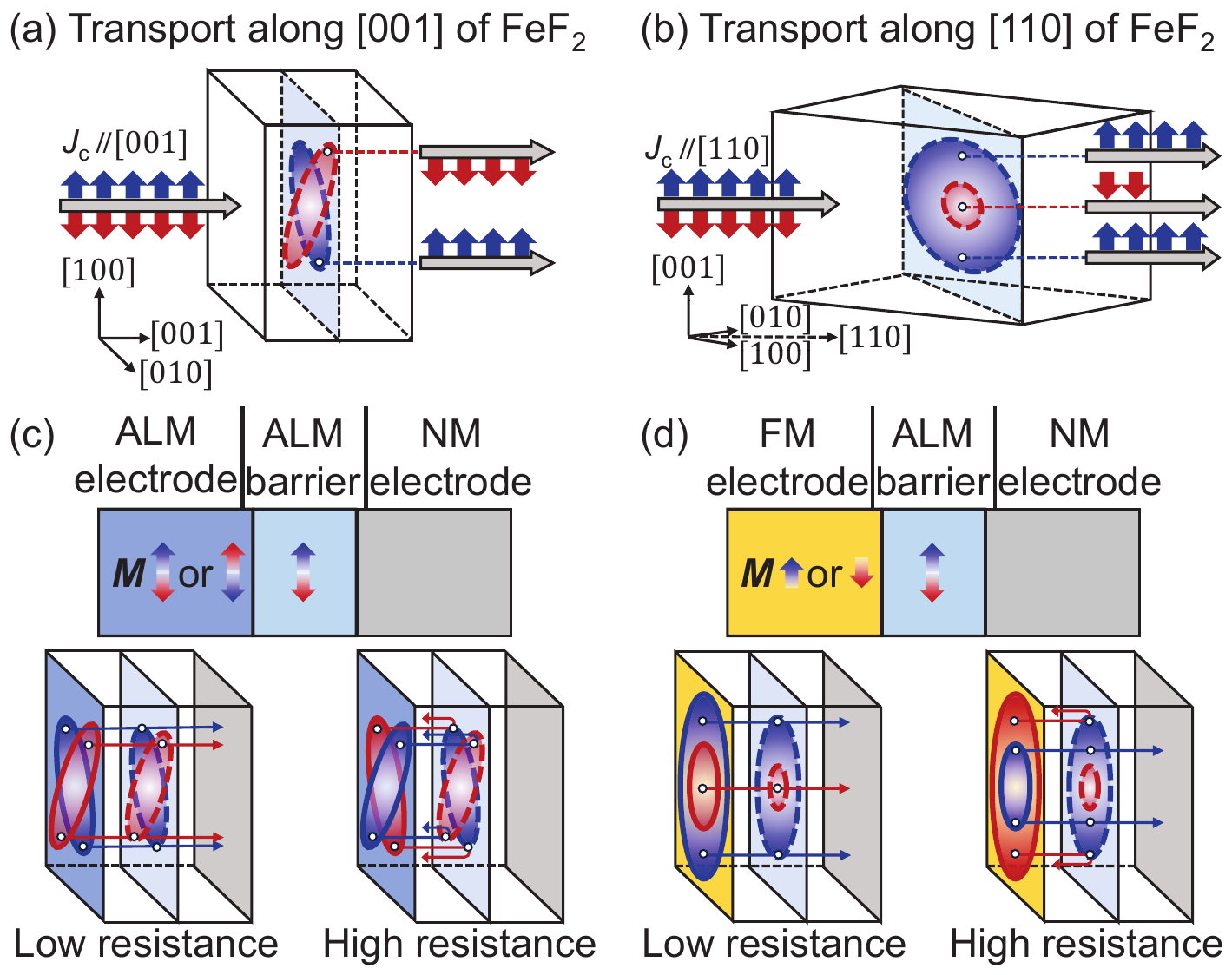}
	\caption{Schematics of the spin filtering effect of ALM insulator FeF$_2$ along (a) the [001] and (b) [110] directions, respectively. Schematics of the TMR mechanism for (c) ALM/ALM insulator barrier/NM tunnel junction with spin-neutral current, (d) FM/ALM insulator barrier/NM tunnel junction with spin-polarized current. Blue and Red circles indicate the spin-up and spin-down channels, respectively. Solid and dashed circles indicate the conduction channels of the electrodes and the area of the evanescent states with low decay rates in the barriers, respectively.}
	\label{fig1}
  \end{figure}

  The properties of ALM materials are closely related to its crystalline orientation \cite{PRA-21-2024,PRB-109-2024}. Electrons undergo different spin filtering effect when tunneling along different crystalline orientations of ALM insulators. Figs. \hyperref[fig1]{1(a)} and \hyperref[fig1]{1(b)} show the crystal-orientation-dependent spin filtering effect when electrons tunnel through the ALM FeF$_2$. Dashed circles denote areas with small decay rates. As shown in Fig. \hyperref[fig1]{1(a)}, when a current flows along the [001] direction of FeF$_2$, due to the same shape of decay rates for the two spin channels, the total spin polarization remains zero, though the current is locally polarized in the momentum space. When a current flows along the [110] direction of FeF$_2$, the current is then globally spin-polarized [Fig. \hyperref[fig1]{1(b)}]. Therefore, the ALM insulators can polarize currents as FM materials, while still retaining advantages of zero stray field and robustness against external magnetic fields.

\begin{figure}
	\centering
	\includegraphics[width=1.0\linewidth]{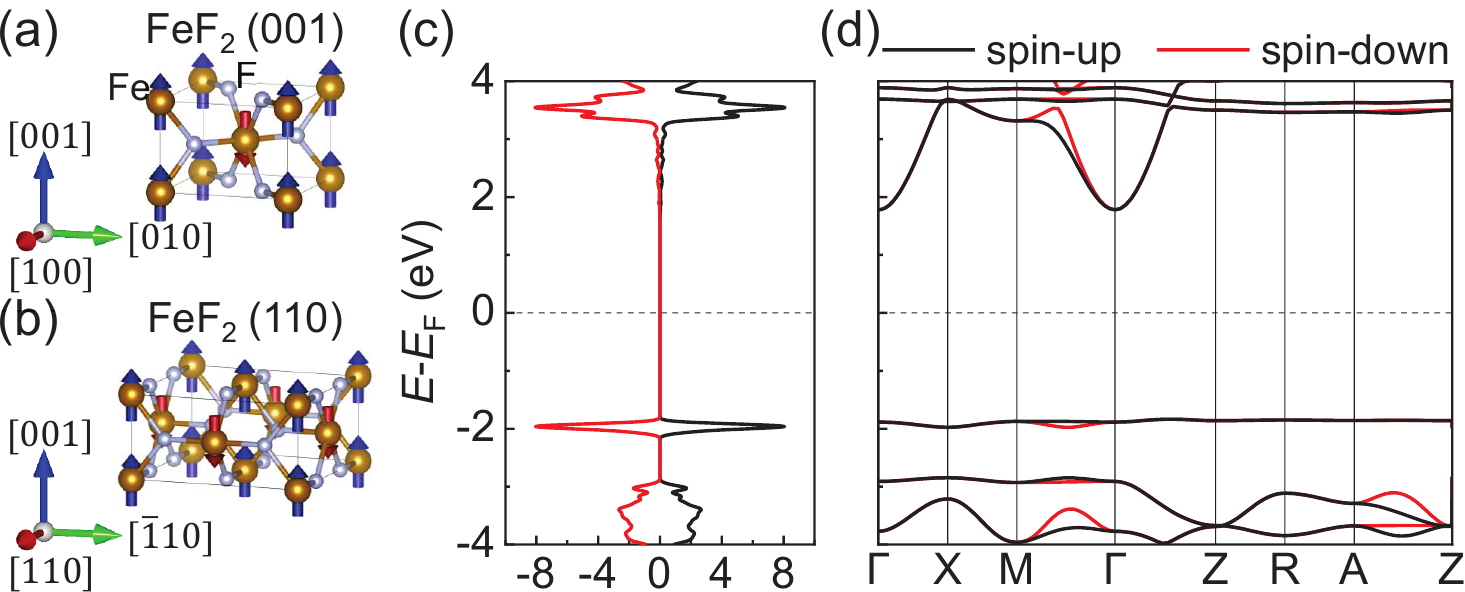}
	\caption{Atomic and magnetic structures of FeF$_2$ with (a) (001) and (b) (110) crystal faces. (c) The density of states and (d) the band structure of FeF$_2$(001).}
	\label{fig2}
  \end{figure}

  To provide a clearer explanation on the spin filtering effect of ALM insulators, we chose FeF$_2$ as an example. FeF$_2$ is an ALM insulator with collinear AFM magnetic configuration and a rutile structure as plotted in Fig. \hyperref[fig2]{2(a)}. Thanks to its magnetic space group $P4_{2}'/mnm’$, FeF$_2$ exhibits a momentum-dependent spin splitting band structure. From the density of states (DOS) of FeF$_2$(001) depicted in Fig. \hyperref[fig2]{2(c)}, FeF$_2$ has zero net magnetization and a 3.5 eV band gap, which is consistent with the experimental study \cite{ACSAMI-2013}. The spin-up and spin-down bands of FeF$_2$(001) in Fig. \hyperref[fig2]{2(d)} degenerate on most high-symmetry lines, but they split along the $\Gamma$-M and Z-A directions, reflecting the altermagnetism. Therefore, when electrons tunnel along the $\Gamma$-M (Z-A) direction, i.e. the [110] direction of FeF$_2$, the band gap of the spin-up and spin-down electrons at this high-symmetry line is diverse [see Supplementary Materials \cite{SM} Fig. S1], which leads to a different barrier height felt by oppositely polarized electrons and thus results in the spin filtering effect.

\begin{figure}
	\centering
	\includegraphics[width=1.0\linewidth]{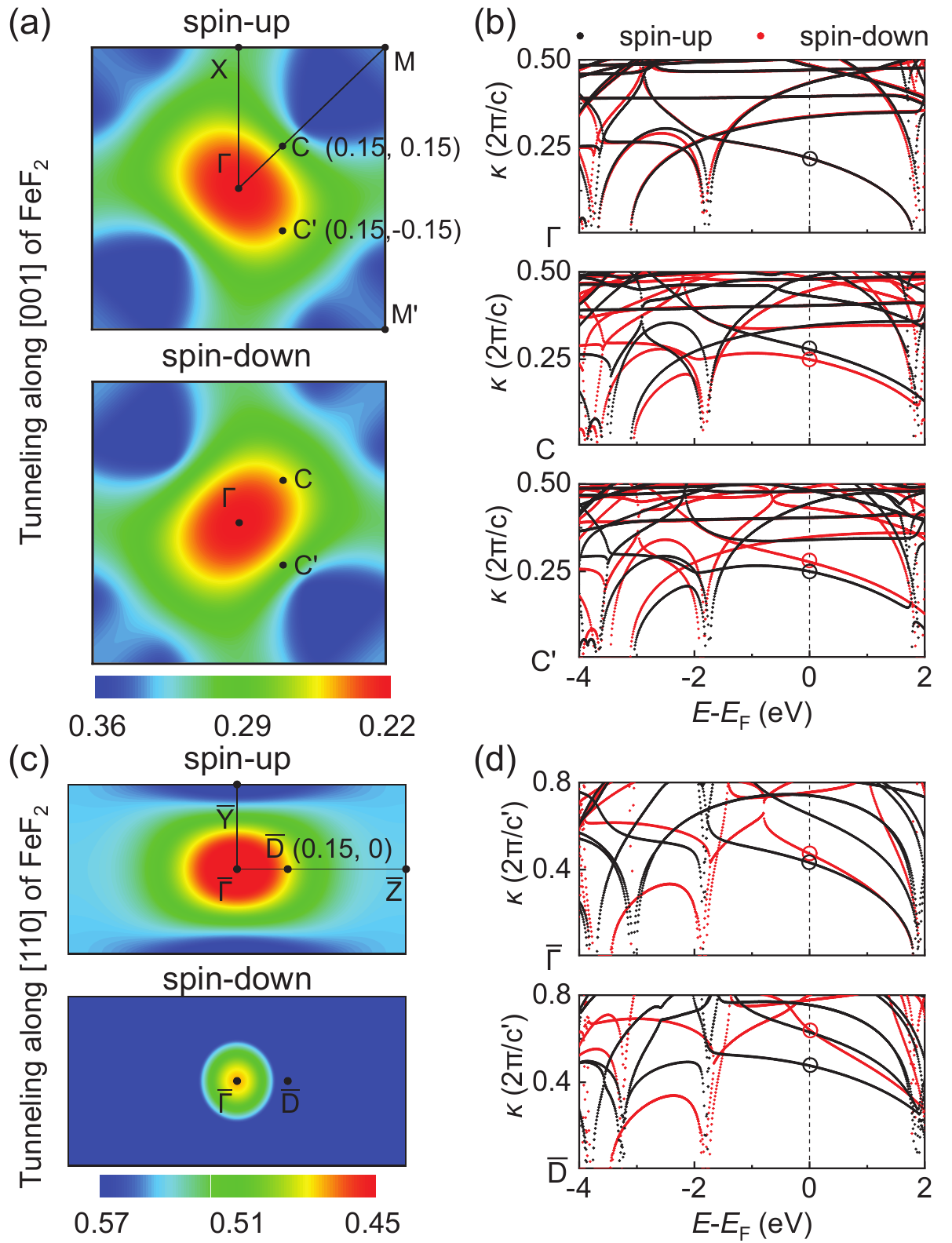}
	\caption{The spin- and $\boldsymbol{k}_\parallel$-resolved lowest decay rates of the evanescent states in FeF$_2$ at the Fermi energy when electrons tunneling along (a) the [001] and (c) the [110] directions. (b) The complex band structures of FeF$_2$ (001) at $\Gamma$, C, C$'$ points. (d) The complex band structures of FeF$_2$ (110) at $\bar{\Gamma}$, $\bar{\rm{D}}$ points. The black and red hollow circles indicate the lowest decay rates for spin-up and spin-down electrons at the Fermi energy, respectively.}
	\label{fig3}
  \end{figure}

  The complex band structure can more accurately illustrate the decay of tunneling electrons inside barrier. In the elastic scattering, the transverse wave vector is conserved, and the longitudinal wave vector $k_z$ can be expressed as $k_z=q+i\kappa$, where the real part $q$ corresponds to the real band, and the imaginary part $\kappa$ corresponds to the wave functions that decay into the bulk \cite{PRL-85-2000}. As a result, tunneling electrons exhibits exponential decay [$\exp(-2\kappa d)$] in the barrier. Figs. \hyperref[fig3]{3(a)} and \hyperref[fig3]{3(b)} show the case of tunneling along the [001] direction of FeF$_2$. The three complex bands in Fig. \hyperref[fig3]{3(b)} correspond to the $\Gamma$(0, 0), C(0.15, 0.15) and C$'$(0.15, -0.15) point. On one hand, for the spin degeneracy at the $\Gamma$ point, the spin-up and spin-down electrons share a similar complex band, thus they have the same lowest decay rate $\kappa$ at the Fermi energy. On the other hand, the spin splitting along $\Gamma$-M leads to a different lowest decay rates for both spins in the complex band structure. However, for FeF$_2$ with altermagnetism, the spin of the $\Gamma$-M$'$ band is opposite to that of the $\Gamma$-M band, resulting in $\kappa_\sigma(\Gamma-\rm{M})=\kappa_{-\sigma}(\Gamma-\rm{M'})$. Eventually, the two spins share the same shape of $\boldsymbol{k}_\parallel$-resolved lowest decay rates, producing a spin-neutral current, but with nonzero local spin polarization in the momentum space.

  Based on this local spin filtering effect in the momentum space, we proposed an MTJ with an ALM metal and an NM metal as electrodes, i.e., ALM/ALM insulator barrier/NM MTJ. As shown in Fig. \hyperref[fig1]{1(c)}, for each spin channel (blue and red), its conductance is proportional to the overlapping between the distribution of the conduction channels in the ALM electrode and the distribution of the lowest decay rates in the barrier. Switching the N\'{e}el vector of the ALM electrode, the spin polarization at each $\boldsymbol{k}_\parallel$ point in the momentum space will be reversed, and the overlapping area between the conduction channels and the lowest decay rates is also changed, resulting in a magnetoresistance between the parallel (P) and antiparallel (AP) states.

  We calculated RuO$_2$(001)/FeF$_2$/IrO$_2$ MTJ as a prototype to verify our design. RuO$_2$ is an ALM metal, and IrO$_2$ is an NM metal, whose conduction channels are shown in Fig. S2. Fig. S3 shows the atomic structure and the transmission coefficient distributions for the P and AP states of this MTJ, it can be seen that the barrier suppresses the transmission coefficients outside the $\Gamma$ point in the AP state, which leads to a TMR ratio of 216\%. This ALM/ALM insulator barrier/NM MTJ can produce the TMR effect with a spin-neutral current and benefit from zero stray field and fast spin dynamics, similar to the AFMTJs \cite{NC-2021,PRL-128-2022,N-613-485-2023,N-613-490-2023,npj2024,arXiv02458}. This [001] transporting MTJ also has the same disadvantages as AFMTJs, which have high demand for single crystals and are hard to be switched by a magnetic field. To overcome these drawbacks, we shift our focus on the transport configuration along the [110] direction.

  Fig. \hyperref[fig2]{2(b)} is the atomic and magnetic structures of FeF$_2$ cleaved from the (110) crystal face. When electrons tunnel along the [110] direction of FeF$_2$, the distribution of the lowest decay rates satisfies Fig. \hyperref[fig3]{3(c)}. Since the band splits along the $\Gamma$-M direction, the two spins have completely different complex band structures at the $\bar{\Gamma}$ point, as shown in Fig. \hyperref[fig3]{3(d)}, and this is also true at other $\boldsymbol{k}_\parallel$ points, resulting in entirely diverse spin-resolved distribution of the lowest decay rates. Consequently, as long as the current flows along the [110] direction, it will be globally polarized, and the spin polarization will rapidly increase with the thickness of FeF$_2$. As a result, the ALM insulator FeF$_2$(110) has the spin filtering effect just like FM insulators.

  Since the spin filtering effect of the FeF$_2$(110) can polarize current globally, the FM electrode is a matched choice of detecting the spin polarization of the current flowing through FeF$_2$(110) by the TMR effect. As depicted in Fig. \hyperref[fig1]{1(d)}, we designed a kind of MTJ composed of FM and NM electrodes, i.e., FM/ALM insulator barrier/NM MTJ. The TMR effect can be generated by switching the magnetization of FM electrode. In this MTJ, only the ALM barrier needs to be single-crystallized for the spin filtering effect. Thus, it is probably easier for experimentalists to grow than the MTJ with both ALM electrode and barrier as discussed before. Furthermore, this MTJ is free from the additional complex pinning structures thanks to the robustness of the ALM reference layer against external fields, and is easily switched by both electrical and magnetic methods, etc.

  \begin{figure}
	\centering
	\includegraphics[width=1.0\linewidth]{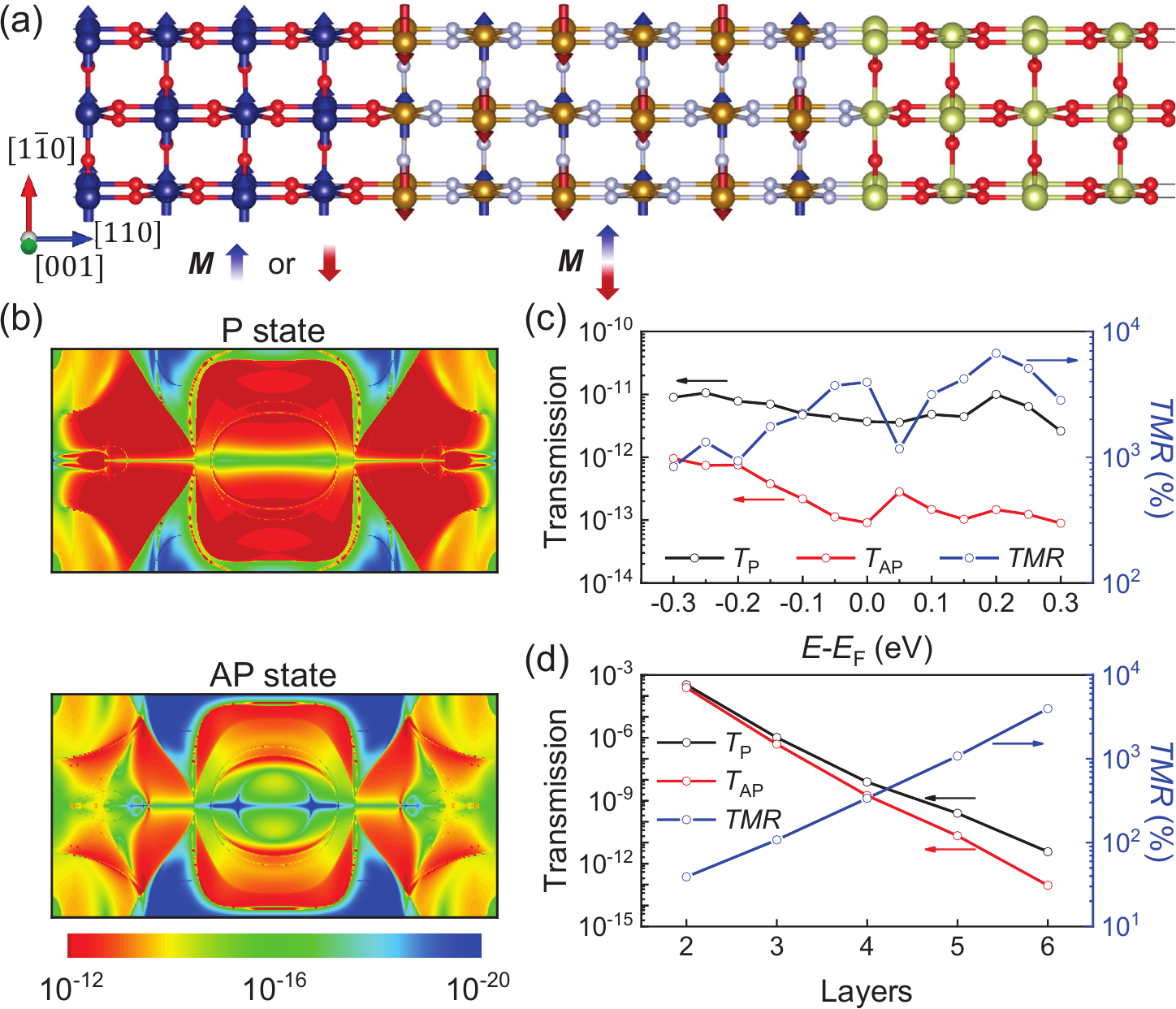}
	\caption{(a) Atomic structure of CrO$_2$(110)/FeF$_2$/IrO$_2$ MTJ. (b) The $\boldsymbol{k}_\parallel$-resolved transmission coefficient distributions in CrO$_2$(110)/FeF$_2$/IrO$_2$ MTJ with P and AP states. Total transmissions and TMR ratios as a function of (c) energy and (d) the number of FeF$_2$ monolayers.}
	\label{fig4}
  \end{figure}

  Hence, we constructed CrO$_2$(110)/FeF$_2$/IrO$_2$ MTJ as an example to exhibit the transport properties of the FM/ALM insulator barrier/NM MTJ. CrO$_2$ is a half-metallic material with 100\% spin polarization, whose conduction channel is shown in Fig. S2(c) of the Supplementary Material \cite{SM}. In fact, CrO$_2$ and IrO$_2$ are only chosen for their same atomic structure with FeF$_2$. We recommend replacing CrO$_2$ and IrO$_2$ with CoFe (or CoFeB etc.) and Pt (or Cu etc.) in experiments, respectively. As shown in Fig. \hyperref[fig4]{4(a)}, the P and AP states are defined as the relative arrangement between the magnetization of CrO$_2$ and the N\'{e}el vector of FeF$_2$. Because CrO$_2$ is a half-metal, there is only spin-up channel in the P state. Switching the magnetization of CrO$_2$, the MTJ becomes the AP state, where only spin-down channel contributes to conductance. For the P state, the higher transmission coefficients appear at the area where the big conduction channels of electrodes match with the smaller $\kappa_\uparrow(\boldsymbol{k}_\parallel)$ of FeF$_2$(110). By comparison, for the AP state, due to the generally larger $\kappa_\downarrow(\boldsymbol{k}_\parallel)$ of FeF$_2$(110), the transmission coefficients are significantly suppressed. Thus, thanks to the large difference of the lowest decay rates of the barrier between the spin-up and spin-down channels, the transmission coefficient of the P state is one order of magnitude larger than that of the AP state, resulting in a huge TMR ratio of 3956\%, which nicely meets the needs for the MRAM and sensor applications.

  Fig. \hyperref[fig4]{4(c)} displays the dependence of total transmissions on energy for the P state ($T_{\rm{P}}$) and AP state ($T_{\rm{AP}}$). It is found that $T_{\rm{P}}$ is always greater than $T_{\rm{AP}}$ around the $E\rm{_F}$, resulting in a positive TMR effect. Specifically, the TMR at the Fermi energy ($E\rm{_F}$) reaches as high as 3956\%. In addition, the TMR ratio remains around 3000\% within the range of $E\rm{_F}$ ± 0.3 eV, indicating that the influence of factors such as doping on the shift of the $E\rm{_F}$ has a small impact on this TMR ratio.

  The spin filtering effect is sensitive to the barrier thickness, as shown in Fig. \hyperref[fig4]{4(d)}, we calculated the total transmissions and TMR ratios as a function of the number of FeF$_2$(110) monolayers. Both $T_{\rm{P}}$ and $T_{\rm{AP}}$ decrease exponentially with increasing barrier layers, while thanks to the spin filtering effect, the TMR ratio increases exponentially, consistent with the simple model in the Supplementary Materials \cite{SM}. As a result, the CrO$_2$(110)/FeF$_2$/IrO$_2$ MTJ can further increase the TMR effect by increasing the barrier thickness, potentially achieving a TMR value significantly greater than 3956\%. Note that this rapid increase in the TMR ratio with barrier thickness does not depend on the selection of FM electrode. Therefore, other FM electrodes, such as CoFeB, Fe, etc., can also be used to replace CrO$_2$ and obtain a large TMR ratio once an appropriate barrier thickness is used. Additionally, since FeF$_2$(110) has the lowest decay rate for electrons with $\Sigma$$_1$ symmetry, selecting an FM electrode where $\Sigma$$_1$ symmetry electrons appear only in the majority state can further enhance the TMR effect by the symmetry filtering effect.
  
  Both proposed ALM barrier-based MTJs have experimental feasibility. To generate the TMR effect, only the ALM layer needs to have good single-crystal properties, while the FM and NM layers are not subject to any restrictions. This makes the FM/ALM barrier/NM MTJ, with only one ALM layer, easier to fabricate. Additionally, since the barrier has AFM configuration, it can serve as the reference layer without additional pinning structures, significantly reducing the complexity of MTJ device fabrication and optimization. Furthermore, the choice of barrier materials for the proposed MTJs is also abundant. The spin filtering effect with AFM configuration is not unique to FeF$_2$, but is an intrinsic property widely shared by ALM insulators, like MnF$_2$ [shown in Supplementary Materials \cite{SM} Fig. S4], MnTe \cite{PRB-107-2023,PRL-132-2024,N-2024}, LaMnO$_3$, etc. \cite{PRX-12-040501-2022}. The MTJ with an ALM barrier constructed based on this effect has a very wide range of candidacy in materials. In our opinion, selecting MnTe with appropriate crystal orientation as the ALM barrier, combined with suitable electrodes, is another good choice for constructing the MTJs proposed in this work, especially considering that the N\'{e}el temperature of MnTe is above the room temperature. Furthermore, if the NM electrode of the FM/ALM insulator barrier/NM MTJ is replaced with an FM electrode, then by changing the magnetization directions of both electrodes and the N\'{e}el vector of the barrier, four different resistance states can be generated, enabling multi-state storage.

  In conclusion, based on the first-principles calculations, we investigated the tunneling process through the ALM insulators. Owing to the spin-momentum locking band structures, the ALM insulators possess spin-resolved lowest decay rates, resulting in an anisotropic spin filtering effect, which has broad application prospects. We demonstrated the crystal-orientation-dependent spin filtering effect in ALM FeF$_2$, and by using RuO$_2$(001)/FeF$_2$/IrO$_2$ and CrO$_2$(110)/FeF$_2$/IrO$_2$ as two prototypes, we proposed two kinds of MTJs, i.e., ALM/ALM insulator barrier/NM and FM/ALM insulator barrier/NM MTJs, and reached 216\% and 3956\% TMR ratios, respectively. The former MTJ benefits from zero stray field, and the latter MTJ has numerous advantages, such as free from an additional complex pinning structure, switchable by both electric methods and magnetic fields, and the only necessity for the single-crystalline structure is the ALM barrier. Thus, these MTJs with an ALM barrier have great potential for applications in the fields of MRAM and magnetic sensors. Our work also provides a promising way to explore the electrical and transport properties of ALM and AFM insulators and promotes the development and application of AFM spintronics.

\begin{acknowledgments}
This work was financially supported by the National Key Research and Development Program of China [MOST Grant No. 2022YFA1402800], the National Natural Science Foundation of China [NSFC, Grant No. 12134017, 12204517, and 12374131], and partially supported by the Strategic Priority Research Program (B) of Chinese Academy of Sciences [CAS Grant No. XDB33000000, Youth Innovation Promotion Association of CAS (2020008)]. The atomic structures were produced using VESTA software \cite{JAC-2011}. High-performance computing resources for contributing to the research results were provided by Beijing PARATERA Technology Co., LTD.
\end{acknowledgments}


\begin{thebibliography}{57}%
	\makeatletter
	\providecommand \@ifxundefined [1]{%
		\@ifx{#1\undefined}
	}%
	\providecommand \@ifnum [1]{%
		\ifnum #1\expandafter \@firstoftwo
		\else \expandafter \@secondoftwo
		\fi
	}%
	\providecommand \@ifx [1]{%
		\ifx #1\expandafter \@firstoftwo
		\else \expandafter \@secondoftwo
		\fi
	}%
	\providecommand \natexlab [1]{#1}%
	\providecommand \enquote  [1]{``#1''}%
	\providecommand \bibnamefont  [1]{#1}%
	\providecommand \bibfnamefont [1]{#1}%
	\providecommand \citenamefont [1]{#1}%
	\providecommand \href@noop [0]{\@secondoftwo}%
	\providecommand \href [0]{\begingroup \@sanitize@url \@href}%
	\providecommand \@href[1]{\@@startlink{#1}\@@href}%
	\providecommand \@@href[1]{\endgroup#1\@@endlink}%
	\providecommand \@sanitize@url [0]{\catcode `\\12\catcode `\$12\catcode
		`\&12\catcode `\#12\catcode `\^12\catcode `\_12\catcode `\%12\relax}%
	\providecommand \@@startlink[1]{}%
	\providecommand \@@endlink[0]{}%
	\providecommand \url  [0]{\begingroup\@sanitize@url \@url }%
	\providecommand \@url [1]{\endgroup\@href {#1}{\urlprefix }}%
	\providecommand \urlprefix  [0]{URL }%
	\providecommand \Eprint [0]{\href }%
	\providecommand \doibase [0]{https://doi.org/}%
	\providecommand \selectlanguage [0]{\@gobble}%
	\providecommand \bibinfo  [0]{\@secondoftwo}%
	\providecommand \bibfield  [0]{\@secondoftwo}%
	\providecommand \translation [1]{[#1]}%
	\providecommand \BibitemOpen [0]{}%
	\providecommand \bibitemStop [0]{}%
	\providecommand \bibitemNoStop [0]{.\EOS\space}%
	\providecommand \EOS [0]{\spacefactor3000\relax}%
	\providecommand \BibitemShut  [1]{\csname bibitem#1\endcsname}%
	\let\auto@bib@innerbib\@empty
	\bibitem [{\citenamefont {Julliere}(1975)}]{PLA-1975}%
	\BibitemOpen
	\bibfield  {author} {\bibinfo {author} {\bibfnamefont {M.}~\bibnamefont
			{Julliere}},\ }\bibfield  {title} {\bibinfo {title} {Tunneling between
			ferromagnetic films},\ }\href
	{https://doi.org/https://doi.org/10.1016/0375-9601(75)90174-7} {\bibfield
		{journal} {\bibinfo  {journal} {Phys. Lett. A}\ }\textbf {\bibinfo {volume}
			{54}},\ \bibinfo {pages} {225} (\bibinfo {year} {1975})}\BibitemShut
	{NoStop}%
	\bibitem [{\citenamefont {Moodera}\ \emph {et~al.}(1995)\citenamefont
		{Moodera}, \citenamefont {Kinder}, \citenamefont {Wong},\ and\ \citenamefont
		{Meservey}}]{PRL-74-1995}%
	\BibitemOpen
	\bibfield  {author} {\bibinfo {author} {\bibfnamefont {J.~S.}\ \bibnamefont
			{Moodera}}, \bibinfo {author} {\bibfnamefont {L.~R.}\ \bibnamefont {Kinder}},
		\bibinfo {author} {\bibfnamefont {T.~M.}\ \bibnamefont {Wong}},\ and\
		\bibinfo {author} {\bibfnamefont {R.}~\bibnamefont {Meservey}},\ }\bibfield
	{title} {\bibinfo {title} {Large magnetoresistance at room temperature in
			ferromagnetic thin film tunnel junctions},\ }\href
	{https://doi.org/10.1103/PhysRevLett.74.3273} {\bibfield  {journal} {\bibinfo
			{journal} {Phys. Rev. Lett.}\ }\textbf {\bibinfo {volume} {74}},\ \bibinfo
		{pages} {3273} (\bibinfo {year} {1995})}\BibitemShut {NoStop}%
	\bibitem [{\citenamefont {Miyazaki}\ and\ \citenamefont
		{Tezuka}(1995)}]{JMMM-1995}%
	\BibitemOpen
	\bibfield  {author} {\bibinfo {author} {\bibfnamefont {T.}~\bibnamefont
			{Miyazaki}}\ and\ \bibinfo {author} {\bibfnamefont {N.}~\bibnamefont
			{Tezuka}},\ }\bibfield  {title} {\bibinfo {title} {Giant magnetic tunneling
			effect in {Fe/Al$_2$O$_3$/Fe} junction},\ }\href
	{https://doi.org/https://doi.org/10.1016/0304-8853(95)90001-2} {\bibfield
		{journal} {\bibinfo  {journal} {J. Magn. Magn. Mater.}\ }\textbf {\bibinfo
			{volume} {139}},\ \bibinfo {pages} {L231} (\bibinfo {year}
		{1995})}\BibitemShut {NoStop}%
	\bibitem [{\citenamefont {Wolf}\ \emph {et~al.}(2001)\citenamefont {Wolf},
		\citenamefont {Awschalom}, \citenamefont {Buhrman}, \citenamefont {Daughton},
		\citenamefont {von Molnár}, \citenamefont {Roukes}, \citenamefont
		{Chtchelkanova},\ and\ \citenamefont {Treger}}]{S-2001}%
	\BibitemOpen
	\bibfield  {author} {\bibinfo {author} {\bibfnamefont {S.~A.}\ \bibnamefont
			{Wolf}}, \bibinfo {author} {\bibfnamefont {D.~D.}\ \bibnamefont {Awschalom}},
		\bibinfo {author} {\bibfnamefont {R.~A.}\ \bibnamefont {Buhrman}}, \bibinfo
		{author} {\bibfnamefont {J.~M.}\ \bibnamefont {Daughton}}, \bibinfo {author}
		{\bibfnamefont {S.}~\bibnamefont {von Molnár}}, \bibinfo {author}
		{\bibfnamefont {M.~L.}\ \bibnamefont {Roukes}}, \bibinfo {author}
		{\bibfnamefont {A.~Y.}\ \bibnamefont {Chtchelkanova}},\ and\ \bibinfo
		{author} {\bibfnamefont {D.~M.}\ \bibnamefont {Treger}},\ }\bibfield  {title}
	{\bibinfo {title} {Spintronics: A spin-based electronics vision for the
			future},\ }\href {https://doi.org/10.1126/science.1065389} {\bibfield
		{journal} {\bibinfo  {journal} {Science}\ }\textbf {\bibinfo {volume}
			{294}},\ \bibinfo {pages} {1488} (\bibinfo {year} {2001})}\BibitemShut
	{NoStop}%
	\bibitem [{\citenamefont {\ifmmode \check{Z}\else
			\v{Z}\fi{}uti\ifmmode~\acute{c}\else \'{c}\fi{}}\ \emph
		{et~al.}(2004)\citenamefont {\ifmmode \check{Z}\else
			\v{Z}\fi{}uti\ifmmode~\acute{c}\else \'{c}\fi{}}, \citenamefont {Fabian},\
		and\ \citenamefont {Das~Sarma}}]{RMP-2004}%
	\BibitemOpen
	\bibfield  {author} {\bibinfo {author} {\bibfnamefont {I.}~\bibnamefont
			{\ifmmode \check{Z}\else \v{Z}\fi{}uti\ifmmode~\acute{c}\else \'{c}\fi{}}},
		\bibinfo {author} {\bibfnamefont {J.}~\bibnamefont {Fabian}},\ and\ \bibinfo
		{author} {\bibfnamefont {S.}~\bibnamefont {Das~Sarma}},\ }\bibfield  {title}
	{\bibinfo {title} {Spintronics: Fundamentals and applications},\ }\href
	{https://doi.org/10.1103/RevModPhys.76.323} {\bibfield  {journal} {\bibinfo
			{journal} {Rev. Mod. Phys.}\ }\textbf {\bibinfo {volume} {76}},\ \bibinfo
		{pages} {323} (\bibinfo {year} {2004})}\BibitemShut {NoStop}%
	\bibitem [{\citenamefont {Lenz}\ and\ \citenamefont
		{Edelstein}(2006)}]{IEEE-2006}%
	\BibitemOpen
	\bibfield  {author} {\bibinfo {author} {\bibfnamefont {J.}~\bibnamefont
			{Lenz}}\ and\ \bibinfo {author} {\bibfnamefont {S.}~\bibnamefont
			{Edelstein}},\ }\bibfield  {title} {\bibinfo {title} {Magnetic sensors and
			their applications},\ }\href {https://doi.org/10.1109/JSEN.2006.874493}
	{\bibfield  {journal} {\bibinfo  {journal} {IEEE Sens. J.}\ }\textbf
		{\bibinfo {volume} {6}},\ \bibinfo {pages} {631} (\bibinfo {year}
		{2006})}\BibitemShut {NoStop}%
	\bibitem [{\citenamefont {Zhang}\ and\ \citenamefont
		{Butler}(2003)}]{JPCM-15-R1603-2003}%
	\BibitemOpen
	\bibfield  {author} {\bibinfo {author} {\bibfnamefont {X.-G.}\ \bibnamefont
			{Zhang}}\ and\ \bibinfo {author} {\bibfnamefont {W.~H.}\ \bibnamefont
			{Butler}},\ }\bibfield  {title} {\bibinfo {title} {Band structure, evanescent
			states, and transport in spin tunnel junctions},\ }\href
	{https://doi.org/10.1088/0953-8984/15/41/R01} {\bibfield  {journal} {\bibinfo
			{journal} {J. Phys.: Condens. Matter}\ }\textbf {\bibinfo {volume} {15}},\
		\bibinfo {pages} {R1603} (\bibinfo {year} {2003})}\BibitemShut {NoStop}%
	\bibitem [{\citenamefont {Tsymbal}\ \emph {et~al.}(2003)\citenamefont
		{Tsymbal}, \citenamefont {Mryasov},\ and\ \citenamefont
		{LeClair}}]{JPCM-15-R109-2003}%
	\BibitemOpen
	\bibfield  {author} {\bibinfo {author} {\bibfnamefont {E.~Y.}\ \bibnamefont
			{Tsymbal}}, \bibinfo {author} {\bibfnamefont {O.~N.}\ \bibnamefont
			{Mryasov}},\ and\ \bibinfo {author} {\bibfnamefont {P.~R.}\ \bibnamefont
			{LeClair}},\ }\bibfield  {title} {\bibinfo {title} {Spin-dependent tunnelling
			in magnetic tunnel junctions},\ }\href
	{https://doi.org/10.1088/0953-8984/15/4/201} {\bibfield  {journal} {\bibinfo
			{journal} {J. Phys.: Condens. Matter}\ }\textbf {\bibinfo {volume} {15}},\
		\bibinfo {pages} {R109} (\bibinfo {year} {2003})}\BibitemShut {NoStop}%
	\bibitem [{\citenamefont {Mavropoulos}\ \emph {et~al.}(2000)\citenamefont
		{Mavropoulos}, \citenamefont {Papanikolaou},\ and\ \citenamefont
		{Dederichs}}]{PRL-85-2000}%
	\BibitemOpen
	\bibfield  {author} {\bibinfo {author} {\bibfnamefont {P.}~\bibnamefont
			{Mavropoulos}}, \bibinfo {author} {\bibfnamefont {N.}~\bibnamefont
			{Papanikolaou}},\ and\ \bibinfo {author} {\bibfnamefont {P.~H.}\ \bibnamefont
			{Dederichs}},\ }\bibfield  {title} {\bibinfo {title} {Complex band structure
			and tunneling through {Ferromagnet}/{Insulator}/{Ferromagnet} junctions},\
	}\href {https://doi.org/10.1103/PhysRevLett.85.1088} {\bibfield  {journal}
		{\bibinfo  {journal} {Phys. Rev. Lett.}\ }\textbf {\bibinfo {volume} {85}},\
		\bibinfo {pages} {1088} (\bibinfo {year} {2000})}\BibitemShut {NoStop}%
	\bibitem [{\citenamefont {Mathon}\ and\ \citenamefont
		{Umerski}(2001)}]{PRB-63-2001-1}%
	\BibitemOpen
	\bibfield  {author} {\bibinfo {author} {\bibfnamefont {J.}~\bibnamefont
			{Mathon}}\ and\ \bibinfo {author} {\bibfnamefont {A.}~\bibnamefont
			{Umerski}},\ }\bibfield  {title} {\bibinfo {title} {Theory of tunneling
			magnetoresistance of an epitaxial {Fe}/{MgO}/{Fe}(001) junction},\ }\href
	{https://doi.org/10.1103/PhysRevB.63.220403} {\bibfield  {journal} {\bibinfo
			{journal} {Phys. Rev. B}\ }\textbf {\bibinfo {volume} {63}},\ \bibinfo
		{pages} {220403} (\bibinfo {year} {2001})}\BibitemShut {NoStop}%
	\bibitem [{\citenamefont {Butler}\ \emph {et~al.}(2001)\citenamefont {Butler},
		\citenamefont {Zhang}, \citenamefont {Schulthess},\ and\ \citenamefont
		{MacLaren}}]{PRB-63-2001-2}%
	\BibitemOpen
	\bibfield  {author} {\bibinfo {author} {\bibfnamefont {W.~H.}\ \bibnamefont
			{Butler}}, \bibinfo {author} {\bibfnamefont {X.-G.}\ \bibnamefont {Zhang}},
		\bibinfo {author} {\bibfnamefont {T.~C.}\ \bibnamefont {Schulthess}},\ and\
		\bibinfo {author} {\bibfnamefont {J.~M.}\ \bibnamefont {MacLaren}},\
	}\bibfield  {title} {\bibinfo {title} {Spin-dependent tunneling conductance
			of {Fe}/{MgO}/{Fe} sandwiches},\ }\href
	{https://doi.org/10.1103/PhysRevB.63.054416} {\bibfield  {journal} {\bibinfo
			{journal} {Phys. Rev. B}\ }\textbf {\bibinfo {volume} {63}},\ \bibinfo
		{pages} {054416} (\bibinfo {year} {2001})}\BibitemShut {NoStop}%
	\bibitem [{\citenamefont {Yuasa}\ \emph {et~al.}(2004)\citenamefont {Yuasa},
		\citenamefont {Nagahama}, \citenamefont {Fukushima}, \citenamefont {Suzuki},\
		and\ \citenamefont {Ando}}]{NM-3-2004-1}%
	\BibitemOpen
	\bibfield  {author} {\bibinfo {author} {\bibfnamefont {S.}~\bibnamefont
			{Yuasa}}, \bibinfo {author} {\bibfnamefont {T.}~\bibnamefont {Nagahama}},
		\bibinfo {author} {\bibfnamefont {A.}~\bibnamefont {Fukushima}}, \bibinfo
		{author} {\bibfnamefont {Y.}~\bibnamefont {Suzuki}},\ and\ \bibinfo {author}
		{\bibfnamefont {K.}~\bibnamefont {Ando}},\ }\bibfield  {title} {\bibinfo
		{title} {Giant room-temperature magnetoresistance in single-crystal
			{Fe}/{MgO}/{Fe} magnetic tunnel junctions},\ }\href
	{https://doi.org/10.1038/nmat1257} {\bibfield  {journal} {\bibinfo  {journal}
			{Nat. Mater.}\ }\textbf {\bibinfo {volume} {3}},\ \bibinfo {pages} {868}
		(\bibinfo {year} {2004})}\BibitemShut {NoStop}%
	\bibitem [{\citenamefont {Parkin}\ \emph {et~al.}(2004)\citenamefont {Parkin},
		\citenamefont {Kaiser}, \citenamefont {Panchula}, \citenamefont {Rice},
		\citenamefont {Hughes}, \citenamefont {Samant},\ and\ \citenamefont
		{Yang}}]{NM-3-2004-2}%
	\BibitemOpen
	\bibfield  {author} {\bibinfo {author} {\bibfnamefont {S.~S.~P.}\
			\bibnamefont {Parkin}}, \bibinfo {author} {\bibfnamefont {C.}~\bibnamefont
			{Kaiser}}, \bibinfo {author} {\bibfnamefont {A.}~\bibnamefont {Panchula}},
		\bibinfo {author} {\bibfnamefont {P.~M.}\ \bibnamefont {Rice}}, \bibinfo
		{author} {\bibfnamefont {B.}~\bibnamefont {Hughes}}, \bibinfo {author}
		{\bibfnamefont {M.}~\bibnamefont {Samant}},\ and\ \bibinfo {author}
		{\bibfnamefont {S.-H.}\ \bibnamefont {Yang}},\ }\bibfield  {title} {\bibinfo
		{title} {Giant tunnelling magnetoresistance at room temperature with {MgO}
			(100) tunnel barriers},\ }\href {https://doi.org/10.1038/nmat1256} {\bibfield
		{journal} {\bibinfo  {journal} {Nat. Mater.}\ }\textbf {\bibinfo {volume}
			{3}},\ \bibinfo {pages} {862} (\bibinfo {year} {2004})}\BibitemShut {NoStop}%
	\bibitem [{\citenamefont {\ifmmode~\check{S}\else \v{S}\fi{}mejkal}\ \emph
		{et~al.}(2022{\natexlab{a}})\citenamefont {\ifmmode~\check{S}\else
			\v{S}\fi{}mejkal}, \citenamefont {Sinova},\ and\ \citenamefont
		{Jungwirth}}]{PRX-12-031042-2022}%
	\BibitemOpen
	\bibfield  {author} {\bibinfo {author} {\bibfnamefont {L.}~\bibnamefont
			{\ifmmode~\check{S}\else \v{S}\fi{}mejkal}}, \bibinfo {author} {\bibfnamefont
			{J.}~\bibnamefont {Sinova}},\ and\ \bibinfo {author} {\bibfnamefont
			{T.}~\bibnamefont {Jungwirth}},\ }\bibfield  {title} {\bibinfo {title}
		{Beyond conventional ferromagnetism and antiferromagnetism: A phase with
			nonrelativistic spin and crystal rotation symmetry},\ }\href
	{https://doi.org/10.1103/PhysRevX.12.031042} {\bibfield  {journal} {\bibinfo
			{journal} {Phys. Rev. X}\ }\textbf {\bibinfo {volume} {12}},\ \bibinfo
		{pages} {031042} (\bibinfo {year} {2022}{\natexlab{a}})}\BibitemShut
	{NoStop}%
	\bibitem [{\citenamefont {\ifmmode~\check{S}\else \v{S}\fi{}mejkal}\ \emph
		{et~al.}(2022{\natexlab{b}})\citenamefont {\ifmmode~\check{S}\else
			\v{S}\fi{}mejkal}, \citenamefont {Sinova},\ and\ \citenamefont
		{Jungwirth}}]{PRX-12-040501-2022}%
	\BibitemOpen
	\bibfield  {author} {\bibinfo {author} {\bibfnamefont {L.}~\bibnamefont
			{\ifmmode~\check{S}\else \v{S}\fi{}mejkal}}, \bibinfo {author} {\bibfnamefont
			{J.}~\bibnamefont {Sinova}},\ and\ \bibinfo {author} {\bibfnamefont
			{T.}~\bibnamefont {Jungwirth}},\ }\bibfield  {title} {\bibinfo {title}
		{Emerging research landscape of altermagnetism},\ }\href
	{https://doi.org/10.1103/PhysRevX.12.040501} {\bibfield  {journal} {\bibinfo
			{journal} {Phys. Rev. X}\ }\textbf {\bibinfo {volume} {12}},\ \bibinfo
		{pages} {040501} (\bibinfo {year} {2022}{\natexlab{b}})}\BibitemShut
	{NoStop}%
	\bibitem [{\citenamefont {Krempask{\'y}}\ \emph {et~al.}(2024)\citenamefont
		{Krempask{\'y}}, \citenamefont {{\v{S}}mejkal}, \citenamefont {D'Souza},
		\citenamefont {Hajlaoui}, \citenamefont {Springholz}, \citenamefont
		{Uhl{\'i}{\v{r}}ov{\'a}}, \citenamefont {Alarab}, \citenamefont
		{Constantinou}, \citenamefont {Strocov}, \citenamefont {Usanov},
		\citenamefont {Pudelko}, \citenamefont {Gonz{\'a}lez-Hern{\'a}ndez},
		\citenamefont {Birk~Hellenes}, \citenamefont {Jansa}, \citenamefont
		{Reichlov{\'a}}, \citenamefont {{\v{S}}ob{\'a}{\v{n}}}, \citenamefont
		{Gonzalez~Betancourt}, \citenamefont {Wadley}, \citenamefont {Sinova},
		\citenamefont {Kriegner}, \citenamefont {Min{\'a}r}, \citenamefont {Dil},\
		and\ \citenamefont {Jungwirth}}]{N-2024}%
	\BibitemOpen
	\bibfield  {author} {\bibinfo {author} {\bibfnamefont {J.}~\bibnamefont
			{Krempask{\'y}}}, \bibinfo {author} {\bibfnamefont {L.}~\bibnamefont
			{{\v{S}}mejkal}}, \bibinfo {author} {\bibfnamefont {S.~W.}\ \bibnamefont
			{D'Souza}}, \bibinfo {author} {\bibfnamefont {M.}~\bibnamefont {Hajlaoui}},
		\bibinfo {author} {\bibfnamefont {G.}~\bibnamefont {Springholz}}, \bibinfo
		{author} {\bibfnamefont {K.}~\bibnamefont {Uhl{\'i}{\v{r}}ov{\'a}}}, \bibinfo
		{author} {\bibfnamefont {F.}~\bibnamefont {Alarab}}, \bibinfo {author}
		{\bibfnamefont {P.~C.}\ \bibnamefont {Constantinou}}, \bibinfo {author}
		{\bibfnamefont {V.}~\bibnamefont {Strocov}}, \bibinfo {author} {\bibfnamefont
			{D.}~\bibnamefont {Usanov}}, \bibinfo {author} {\bibfnamefont {W.~R.}\
			\bibnamefont {Pudelko}}, \bibinfo {author} {\bibfnamefont {R.}~\bibnamefont
			{Gonz{\'a}lez-Hern{\'a}ndez}}, \bibinfo {author} {\bibfnamefont
			{A.}~\bibnamefont {Birk~Hellenes}}, \bibinfo {author} {\bibfnamefont
			{Z.}~\bibnamefont {Jansa}}, \bibinfo {author} {\bibfnamefont
			{H.}~\bibnamefont {Reichlov{\'a}}}, \bibinfo {author} {\bibfnamefont
			{Z.}~\bibnamefont {{\v{S}}ob{\'a}{\v{n}}}}, \bibinfo {author} {\bibfnamefont
			{R.~D.}\ \bibnamefont {Gonzalez~Betancourt}}, \bibinfo {author}
		{\bibfnamefont {P.}~\bibnamefont {Wadley}}, \bibinfo {author} {\bibfnamefont
			{J.}~\bibnamefont {Sinova}}, \bibinfo {author} {\bibfnamefont
			{D.}~\bibnamefont {Kriegner}}, \bibinfo {author} {\bibfnamefont
			{J.}~\bibnamefont {Min{\'a}r}}, \bibinfo {author} {\bibfnamefont {J.~H.}\
			\bibnamefont {Dil}},\ and\ \bibinfo {author} {\bibfnamefont {T.}~\bibnamefont
			{Jungwirth}},\ }\bibfield  {title} {\bibinfo {title} {Altermagnetic lifting
			of kramers spin degeneracy},\ }\href
	{https://doi.org/10.1038/s41586-023-06907-7} {\bibfield  {journal} {\bibinfo
			{journal} {Nature}\ }\textbf {\bibinfo {volume} {626}},\ \bibinfo {pages}
		{517} (\bibinfo {year} {2024})}\BibitemShut {NoStop}%
	\bibitem [{\citenamefont {Shao}\ and\ \citenamefont {Tsymbal}(2024)}]{npj2024}%
	\BibitemOpen
	\bibfield  {author} {\bibinfo {author} {\bibfnamefont {D.-F.}\ \bibnamefont
			{Shao}}\ and\ \bibinfo {author} {\bibfnamefont {E.~Y.}\ \bibnamefont
			{Tsymbal}},\ }\bibfield  {title} {\bibinfo {title} {Antiferromagnetic tunnel
			junctions for spintronics},\ }\href
	{https://doi.org/10.1038/s44306-024-00014-7} {\bibfield  {journal} {\bibinfo
			{journal} {npj Spintronics}\ }\textbf {\bibinfo {volume} {2}},\ \bibinfo
		{pages} {13} (\bibinfo {year} {2024})}\BibitemShut {NoStop}%
	\bibitem [{\citenamefont {Shao}\ \emph {et~al.}(2023)\citenamefont {Shao},
		\citenamefont {Jiang}, \citenamefont {Ding}, \citenamefont {Zhang},
		\citenamefont {Wang}, \citenamefont {Xiao}, \citenamefont {Gurung},
		\citenamefont {Lu}, \citenamefont {Sun},\ and\ \citenamefont
		{Tsymbal}}]{PRL-130-2023}%
	\BibitemOpen
	\bibfield  {author} {\bibinfo {author} {\bibfnamefont {D.-F.}\ \bibnamefont
			{Shao}}, \bibinfo {author} {\bibfnamefont {Y.-Y.}\ \bibnamefont {Jiang}},
		\bibinfo {author} {\bibfnamefont {J.}~\bibnamefont {Ding}}, \bibinfo {author}
		{\bibfnamefont {S.-H.}\ \bibnamefont {Zhang}}, \bibinfo {author}
		{\bibfnamefont {Z.-A.}\ \bibnamefont {Wang}}, \bibinfo {author}
		{\bibfnamefont {R.-C.}\ \bibnamefont {Xiao}}, \bibinfo {author}
		{\bibfnamefont {G.}~\bibnamefont {Gurung}}, \bibinfo {author} {\bibfnamefont
			{W.~J.}\ \bibnamefont {Lu}}, \bibinfo {author} {\bibfnamefont {Y.~P.}\
			\bibnamefont {Sun}},\ and\ \bibinfo {author} {\bibfnamefont {E.~Y.}\
			\bibnamefont {Tsymbal}},\ }\bibfield  {title} {\bibinfo {title} {N\'eel spin
			currents in antiferromagnets},\ }\href
	{https://doi.org/10.1103/PhysRevLett.130.216702} {\bibfield  {journal}
		{\bibinfo  {journal} {Phys. Rev. Lett.}\ }\textbf {\bibinfo {volume} {130}},\
		\bibinfo {pages} {216702} (\bibinfo {year} {2023})}\BibitemShut {NoStop}%
	\bibitem [{\citenamefont {Leivisk\"a}\ \emph {et~al.}(2024)\citenamefont
		{Leivisk\"a}, \citenamefont {Rial}, \citenamefont {Bad'ura}, \citenamefont
		{Seeger}, \citenamefont {Kounta}, \citenamefont {Beckert}, \citenamefont
		{Kriegner}, \citenamefont {Joumard}, \citenamefont {Schmoranzerov\'a},
		\citenamefont {Sinova}, \citenamefont {Gomonay}, \citenamefont {Thomas},
		\citenamefont {Goennenwein}, \citenamefont {Reichlov\'a}, \citenamefont
		{\ifmmode~\check{S}\else \v{S}\fi{}mejkal}, \citenamefont {Michez},
		\citenamefont {Jungwirth},\ and\ \citenamefont
		{Baltz}}]{PRB-109-2024-224430}%
	\BibitemOpen
	\bibfield  {author} {\bibinfo {author} {\bibfnamefont {M.}~\bibnamefont
			{Leivisk\"a}}, \bibinfo {author} {\bibfnamefont {J.}~\bibnamefont {Rial}},
		\bibinfo {author} {\bibfnamefont {A.}~\bibnamefont {Bad'ura}}, \bibinfo
		{author} {\bibfnamefont {R.~L.}\ \bibnamefont {Seeger}}, \bibinfo {author}
		{\bibfnamefont {I.}~\bibnamefont {Kounta}}, \bibinfo {author} {\bibfnamefont
			{S.}~\bibnamefont {Beckert}}, \bibinfo {author} {\bibfnamefont
			{D.}~\bibnamefont {Kriegner}}, \bibinfo {author} {\bibfnamefont
			{I.}~\bibnamefont {Joumard}}, \bibinfo {author} {\bibfnamefont
			{E.}~\bibnamefont {Schmoranzerov\'a}}, \bibinfo {author} {\bibfnamefont
			{J.}~\bibnamefont {Sinova}}, \bibinfo {author} {\bibfnamefont
			{O.}~\bibnamefont {Gomonay}}, \bibinfo {author} {\bibfnamefont
			{A.}~\bibnamefont {Thomas}}, \bibinfo {author} {\bibfnamefont {S.~T.~B.}\
			\bibnamefont {Goennenwein}}, \bibinfo {author} {\bibfnamefont
			{H.}~\bibnamefont {Reichlov\'a}}, \bibinfo {author} {\bibfnamefont
			{L.}~\bibnamefont {\ifmmode~\check{S}\else \v{S}\fi{}mejkal}}, \bibinfo
		{author} {\bibfnamefont {L.}~\bibnamefont {Michez}}, \bibinfo {author}
		{\bibfnamefont {T.~c.~v.}\ \bibnamefont {Jungwirth}},\ and\ \bibinfo {author}
		{\bibfnamefont {V.}~\bibnamefont {Baltz}},\ }\bibfield  {title} {\bibinfo
		{title} {{Anisotropy of the anomalous Hall effect in thin films of the
				altermagnet candidate ${\mathrm{Mn}}_{5}{\mathrm{Si}}_{3}$}},\ }\href
	{https://doi.org/10.1103/PhysRevB.109.224430} {\bibfield  {journal} {\bibinfo
			{journal} {Phys. Rev. B}\ }\textbf {\bibinfo {volume} {109}},\ \bibinfo
		{pages} {224430} (\bibinfo {year} {2024})}\BibitemShut {NoStop}%
	\bibitem [{\citenamefont {Mazin}(2023)}]{PRB-107-2023}%
	\BibitemOpen
	\bibfield  {author} {\bibinfo {author} {\bibfnamefont {I.~I.}\ \bibnamefont
			{Mazin}},\ }\bibfield  {title} {\bibinfo {title} {Altermagnetism in {MnTe}:
			Origin, predicted manifestations, and routes to detwinning},\ }\href
	{https://doi.org/10.1103/PhysRevB.107.L100418} {\bibfield  {journal}
		{\bibinfo  {journal} {Phys. Rev. B}\ }\textbf {\bibinfo {volume} {107}},\
		\bibinfo {pages} {L100418} (\bibinfo {year} {2023})}\BibitemShut {NoStop}%
	\bibitem [{\citenamefont {Lee}\ \emph {et~al.}(2024)\citenamefont {Lee},
		\citenamefont {Lee}, \citenamefont {Jung}, \citenamefont {Jung},
		\citenamefont {Kim}, \citenamefont {Lee}, \citenamefont {Seok}, \citenamefont
		{Kim}, \citenamefont {Park}, \citenamefont {\ifmmode~\check{S}\else
			\v{S}\fi{}mejkal}, \citenamefont {Kang},\ and\ \citenamefont
		{Kim}}]{PRL-132-2024}%
	\BibitemOpen
	\bibfield  {author} {\bibinfo {author} {\bibfnamefont {S.}~\bibnamefont
			{Lee}}, \bibinfo {author} {\bibfnamefont {S.}~\bibnamefont {Lee}}, \bibinfo
		{author} {\bibfnamefont {S.}~\bibnamefont {Jung}}, \bibinfo {author}
		{\bibfnamefont {J.}~\bibnamefont {Jung}}, \bibinfo {author} {\bibfnamefont
			{D.}~\bibnamefont {Kim}}, \bibinfo {author} {\bibfnamefont {Y.}~\bibnamefont
			{Lee}}, \bibinfo {author} {\bibfnamefont {B.}~\bibnamefont {Seok}}, \bibinfo
		{author} {\bibfnamefont {J.}~\bibnamefont {Kim}}, \bibinfo {author}
		{\bibfnamefont {B.~G.}\ \bibnamefont {Park}}, \bibinfo {author}
		{\bibfnamefont {L.}~\bibnamefont {\ifmmode~\check{S}\else \v{S}\fi{}mejkal}},
		\bibinfo {author} {\bibfnamefont {C.-J.}\ \bibnamefont {Kang}},\ and\
		\bibinfo {author} {\bibfnamefont {C.}~\bibnamefont {Kim}},\ }\bibfield
	{title} {\bibinfo {title} {Broken kramers degeneracy in altermagnetic mnte},\
	}\href {https://doi.org/10.1103/PhysRevLett.132.036702} {\bibfield  {journal}
		{\bibinfo  {journal} {Phys. Rev. Lett.}\ }\textbf {\bibinfo {volume} {132}},\
		\bibinfo {pages} {036702} (\bibinfo {year} {2024})}\BibitemShut {NoStop}%
	\bibitem [{\citenamefont {Banerjee}\ and\ \citenamefont
		{Scheurer}(2024)}]{PRB-110-2024-024503}%
	\BibitemOpen
	\bibfield  {author} {\bibinfo {author} {\bibfnamefont {S.}~\bibnamefont
			{Banerjee}}\ and\ \bibinfo {author} {\bibfnamefont {M.~S.}\ \bibnamefont
			{Scheurer}},\ }\bibfield  {title} {\bibinfo {title} {Altermagnetic
			superconducting diode effect},\ }\href
	{https://doi.org/10.1103/PhysRevB.110.024503} {\bibfield  {journal} {\bibinfo
			{journal} {Phys. Rev. B}\ }\textbf {\bibinfo {volume} {110}},\ \bibinfo
		{pages} {024503} (\bibinfo {year} {2024})}\BibitemShut {NoStop}%
	\bibitem [{\citenamefont {Antonenko}\ \emph {et~al.}(2024)\citenamefont
		{Antonenko}, \citenamefont {Fernandes},\ and\ \citenamefont
		{Venderbos}}]{arXiv10201}%
	\BibitemOpen
	\bibfield  {author} {\bibinfo {author} {\bibfnamefont {D.~S.}\ \bibnamefont
			{Antonenko}}, \bibinfo {author} {\bibfnamefont {R.~M.}\ \bibnamefont
			{Fernandes}},\ and\ \bibinfo {author} {\bibfnamefont {J.~W.~F.}\ \bibnamefont
			{Venderbos}},\ }\href@noop {} {\bibinfo {title} {Mirror chern bands and weyl
			nodal loops in altermagnets}} (\bibinfo {year} {2024}),\ \Eprint
	{https://arxiv.org/abs/2402.10201} {arXiv:2402.10201} \BibitemShut {NoStop}%
	\bibitem [{\citenamefont {Das}\ and\ \citenamefont {Roy}(2024)}]{arXiv14620}%
	\BibitemOpen
	\bibfield  {author} {\bibinfo {author} {\bibfnamefont {S.~K.}\ \bibnamefont
			{Das}}\ and\ \bibinfo {author} {\bibfnamefont {B.}~\bibnamefont {Roy}},\
	}\href@noop {} {\bibinfo {title} {From local to emergent altermagnetism:
			Footprints of free fermions band topology}} (\bibinfo {year} {2024}),\
	\Eprint {https://arxiv.org/abs/2403.14620} {arXiv:2403.14620} \BibitemShut
	{NoStop}%
	\bibitem [{\citenamefont {Zhu}\ \emph {et~al.}(2024)\citenamefont {Zhu},
		\citenamefont {Chen}, \citenamefont {Li}, \citenamefont {Qiao}, \citenamefont
		{Ma}, \citenamefont {Liu}, \citenamefont {Hu}, \citenamefont {Gao},\ and\
		\citenamefont {Ren}}]{NL-2024}%
	\BibitemOpen
	\bibfield  {author} {\bibinfo {author} {\bibfnamefont {Y.}~\bibnamefont
			{Zhu}}, \bibinfo {author} {\bibfnamefont {T.}~\bibnamefont {Chen}}, \bibinfo
		{author} {\bibfnamefont {Y.}~\bibnamefont {Li}}, \bibinfo {author}
		{\bibfnamefont {L.}~\bibnamefont {Qiao}}, \bibinfo {author} {\bibfnamefont
			{X.}~\bibnamefont {Ma}}, \bibinfo {author} {\bibfnamefont {C.}~\bibnamefont
			{Liu}}, \bibinfo {author} {\bibfnamefont {T.}~\bibnamefont {Hu}}, \bibinfo
		{author} {\bibfnamefont {H.}~\bibnamefont {Gao}},\ and\ \bibinfo {author}
		{\bibfnamefont {W.}~\bibnamefont {Ren}},\ }\bibfield  {title} {\bibinfo
		{title} {Multipiezo effect in altermagnetic {V$_2$SeTeO} monolayer},\ }\href
	{https://doi.org/10.1021/acs.nanolett.3c04330} {\bibfield  {journal}
		{\bibinfo  {journal} {Nano Lett.}\ }\textbf {\bibinfo {volume} {24}},\
		\bibinfo {pages} {472} (\bibinfo {year} {2024})}\BibitemShut {NoStop}%
	\bibitem [{\citenamefont {Shao}\ \emph {et~al.}(2021)\citenamefont {Shao},
		\citenamefont {Zhang}, \citenamefont {Li}, \citenamefont {Eom},\ and\
		\citenamefont {Tsymbal}}]{NC-2021}%
	\BibitemOpen
	\bibfield  {author} {\bibinfo {author} {\bibfnamefont {D.-F.}\ \bibnamefont
			{Shao}}, \bibinfo {author} {\bibfnamefont {S.-H.}\ \bibnamefont {Zhang}},
		\bibinfo {author} {\bibfnamefont {M.}~\bibnamefont {Li}}, \bibinfo {author}
		{\bibfnamefont {C.-B.}\ \bibnamefont {Eom}},\ and\ \bibinfo {author}
		{\bibfnamefont {E.~Y.}\ \bibnamefont {Tsymbal}},\ }\bibfield  {title}
	{\bibinfo {title} {Spin-neutral currents for spintronics},\ }\href
	{https://doi.org/10.1038/s41467-021-26915-3} {\bibfield  {journal} {\bibinfo
			{journal} {Nat. Commun.}\ }\textbf {\bibinfo {volume} {12}},\ \bibinfo
		{pages} {7061} (\bibinfo {year} {2021})}\BibitemShut {NoStop}%
	\bibitem [{\citenamefont {Dong}\ \emph {et~al.}(2022)\citenamefont {Dong},
		\citenamefont {Li}, \citenamefont {Gurung}, \citenamefont {Zhu},
		\citenamefont {Zhang}, \citenamefont {Zheng}, \citenamefont {Tsymbal},\ and\
		\citenamefont {Zhang}}]{PRL-128-2022}%
	\BibitemOpen
	\bibfield  {author} {\bibinfo {author} {\bibfnamefont {J.}~\bibnamefont
			{Dong}}, \bibinfo {author} {\bibfnamefont {X.}~\bibnamefont {Li}}, \bibinfo
		{author} {\bibfnamefont {G.}~\bibnamefont {Gurung}}, \bibinfo {author}
		{\bibfnamefont {M.}~\bibnamefont {Zhu}}, \bibinfo {author} {\bibfnamefont
			{P.}~\bibnamefont {Zhang}}, \bibinfo {author} {\bibfnamefont
			{F.}~\bibnamefont {Zheng}}, \bibinfo {author} {\bibfnamefont {E.~Y.}\
			\bibnamefont {Tsymbal}},\ and\ \bibinfo {author} {\bibfnamefont
			{J.}~\bibnamefont {Zhang}},\ }\bibfield  {title} {\bibinfo {title} {Tunneling
			magnetoresistance in noncollinear antiferromagnetic tunnel junctions},\
	}\href {https://doi.org/10.1103/PhysRevLett.128.197201} {\bibfield  {journal}
		{\bibinfo  {journal} {Phys. Rev. Lett.}\ }\textbf {\bibinfo {volume} {128}},\
		\bibinfo {pages} {197201} (\bibinfo {year} {2022})}\BibitemShut {NoStop}%
	\bibitem [{\citenamefont {Qin}\ \emph {et~al.}(2023)\citenamefont {Qin},
		\citenamefont {Yan}, \citenamefont {Wang}, \citenamefont {Chen},
		\citenamefont {Meng}, \citenamefont {Dong}, \citenamefont {Zhu},
		\citenamefont {Cai}, \citenamefont {Feng}, \citenamefont {Zhou},
		\citenamefont {Liu}, \citenamefont {Zhang}, \citenamefont {Zeng},
		\citenamefont {Zhang}, \citenamefont {Jiang},\ and\ \citenamefont
		{Liu}}]{N-613-485-2023}%
	\BibitemOpen
	\bibfield  {author} {\bibinfo {author} {\bibfnamefont {P.}~\bibnamefont
			{Qin}}, \bibinfo {author} {\bibfnamefont {H.}~\bibnamefont {Yan}}, \bibinfo
		{author} {\bibfnamefont {X.}~\bibnamefont {Wang}}, \bibinfo {author}
		{\bibfnamefont {H.}~\bibnamefont {Chen}}, \bibinfo {author} {\bibfnamefont
			{Z.}~\bibnamefont {Meng}}, \bibinfo {author} {\bibfnamefont {J.}~\bibnamefont
			{Dong}}, \bibinfo {author} {\bibfnamefont {M.}~\bibnamefont {Zhu}}, \bibinfo
		{author} {\bibfnamefont {J.}~\bibnamefont {Cai}}, \bibinfo {author}
		{\bibfnamefont {Z.}~\bibnamefont {Feng}}, \bibinfo {author} {\bibfnamefont
			{X.}~\bibnamefont {Zhou}}, \bibinfo {author} {\bibfnamefont {L.}~\bibnamefont
			{Liu}}, \bibinfo {author} {\bibfnamefont {T.}~\bibnamefont {Zhang}}, \bibinfo
		{author} {\bibfnamefont {Z.}~\bibnamefont {Zeng}}, \bibinfo {author}
		{\bibfnamefont {J.}~\bibnamefont {Zhang}}, \bibinfo {author} {\bibfnamefont
			{C.}~\bibnamefont {Jiang}},\ and\ \bibinfo {author} {\bibfnamefont
			{Z.}~\bibnamefont {Liu}},\ }\bibfield  {title} {\bibinfo {title}
		{Room-temperature magnetoresistance in an all-antiferromagnetic tunnel
			junction},\ }\href {https://doi.org/10.1038/s41586-022-05461-y} {\bibfield
		{journal} {\bibinfo  {journal} {Nature}\ }\textbf {\bibinfo {volume} {613}},\
		\bibinfo {pages} {485} (\bibinfo {year} {2023})}\BibitemShut {NoStop}%
	\bibitem [{\citenamefont {Chen}\ \emph {et~al.}(2023)\citenamefont {Chen},
		\citenamefont {Higo}, \citenamefont {Tanaka}, \citenamefont {Nomoto},
		\citenamefont {Tsai}, \citenamefont {Idzuchi}, \citenamefont {Shiga},
		\citenamefont {Sakamoto}, \citenamefont {Ando}, \citenamefont {Kosaki},
		\citenamefont {Matsuo}, \citenamefont {Nishio-Hamane}, \citenamefont {Arita},
		\citenamefont {Miwa},\ and\ \citenamefont {Nakatsuji}}]{N-613-490-2023}%
	\BibitemOpen
	\bibfield  {author} {\bibinfo {author} {\bibfnamefont {X.}~\bibnamefont
			{Chen}}, \bibinfo {author} {\bibfnamefont {T.}~\bibnamefont {Higo}}, \bibinfo
		{author} {\bibfnamefont {K.}~\bibnamefont {Tanaka}}, \bibinfo {author}
		{\bibfnamefont {T.}~\bibnamefont {Nomoto}}, \bibinfo {author} {\bibfnamefont
			{H.}~\bibnamefont {Tsai}}, \bibinfo {author} {\bibfnamefont {H.}~\bibnamefont
			{Idzuchi}}, \bibinfo {author} {\bibfnamefont {M.}~\bibnamefont {Shiga}},
		\bibinfo {author} {\bibfnamefont {S.}~\bibnamefont {Sakamoto}}, \bibinfo
		{author} {\bibfnamefont {R.}~\bibnamefont {Ando}}, \bibinfo {author}
		{\bibfnamefont {H.}~\bibnamefont {Kosaki}}, \bibinfo {author} {\bibfnamefont
			{T.}~\bibnamefont {Matsuo}}, \bibinfo {author} {\bibfnamefont
			{D.}~\bibnamefont {Nishio-Hamane}}, \bibinfo {author} {\bibfnamefont
			{R.}~\bibnamefont {Arita}}, \bibinfo {author} {\bibfnamefont
			{S.}~\bibnamefont {Miwa}},\ and\ \bibinfo {author} {\bibfnamefont
			{S.}~\bibnamefont {Nakatsuji}},\ }\bibfield  {title} {\bibinfo {title}
		{Octupole-driven magnetoresistance in an antiferromagnetic tunnel junction},\
	}\href {https://doi.org/10.1038/s41586-022-05463-w} {\bibfield  {journal}
		{\bibinfo  {journal} {Nature}\ }\textbf {\bibinfo {volume} {613}},\ \bibinfo
		{pages} {490} (\bibinfo {year} {2023})}\BibitemShut {NoStop}%
	\bibitem [{\citenamefont {Jiang}\ \emph {et~al.}(2023)\citenamefont {Jiang},
		\citenamefont {Wang}, \citenamefont {Samanta}, \citenamefont {Zhang},
		\citenamefont {Xiao}, \citenamefont {Lu}, \citenamefont {Sun}, \citenamefont
		{Tsymbal},\ and\ \citenamefont {Shao}}]{PRB-108-2023}%
	\BibitemOpen
	\bibfield  {author} {\bibinfo {author} {\bibfnamefont {Y.-Y.}\ \bibnamefont
			{Jiang}}, \bibinfo {author} {\bibfnamefont {Z.-A.}\ \bibnamefont {Wang}},
		\bibinfo {author} {\bibfnamefont {K.}~\bibnamefont {Samanta}}, \bibinfo
		{author} {\bibfnamefont {S.-H.}\ \bibnamefont {Zhang}}, \bibinfo {author}
		{\bibfnamefont {R.-C.}\ \bibnamefont {Xiao}}, \bibinfo {author}
		{\bibfnamefont {W.~J.}\ \bibnamefont {Lu}}, \bibinfo {author} {\bibfnamefont
			{Y.~P.}\ \bibnamefont {Sun}}, \bibinfo {author} {\bibfnamefont {E.~Y.}\
			\bibnamefont {Tsymbal}},\ and\ \bibinfo {author} {\bibfnamefont {D.-F.}\
			\bibnamefont {Shao}},\ }\bibfield  {title} {\bibinfo {title} {Prediction of
			giant tunneling magnetoresistance in {RuO$_2$}/{TiO$_2$}/{RuO$_2$} (110)
			antiferromagnetic tunnel junctions},\ }\href
	{https://doi.org/10.1103/PhysRevB.108.174439} {\bibfield  {journal} {\bibinfo
			{journal} {Phys. Rev. B}\ }\textbf {\bibinfo {volume} {108}},\ \bibinfo
		{pages} {174439} (\bibinfo {year} {2023})}\BibitemShut {NoStop}%
	\bibitem [{\citenamefont {\ifmmode~\check{S}\else \v{S}\fi{}mejkal}\ \emph
		{et~al.}(2022{\natexlab{c}})\citenamefont {\ifmmode~\check{S}\else
			\v{S}\fi{}mejkal}, \citenamefont {Hellenes}, \citenamefont
		{Gonz\'alez-Hern\'andez}, \citenamefont {Sinova},\ and\ \citenamefont
		{Jungwirth}}]{PRX-12-011028-2022}%
	\BibitemOpen
	\bibfield  {author} {\bibinfo {author} {\bibfnamefont {L.}~\bibnamefont
			{\ifmmode~\check{S}\else \v{S}\fi{}mejkal}}, \bibinfo {author} {\bibfnamefont
			{A.~B.}\ \bibnamefont {Hellenes}}, \bibinfo {author} {\bibfnamefont
			{R.}~\bibnamefont {Gonz\'alez-Hern\'andez}}, \bibinfo {author} {\bibfnamefont
			{J.}~\bibnamefont {Sinova}},\ and\ \bibinfo {author} {\bibfnamefont
			{T.}~\bibnamefont {Jungwirth}},\ }\bibfield  {title} {\bibinfo {title} {Giant
			and tunneling magnetoresistance in unconventional collinear antiferromagnets
			with nonrelativistic spin-momentum coupling},\ }\href
	{https://doi.org/10.1103/PhysRevX.12.011028} {\bibfield  {journal} {\bibinfo
			{journal} {Phys. Rev. X}\ }\textbf {\bibinfo {volume} {12}},\ \bibinfo
		{pages} {011028} (\bibinfo {year} {2022}{\natexlab{c}})}\BibitemShut
	{NoStop}%
	\bibitem [{\citenamefont {Xu}\ \emph {et~al.}(2023)\citenamefont {Xu},
		\citenamefont {Zhang}, \citenamefont {Mahfouzi}, \citenamefont {Huang},
		\citenamefont {Cheng}, \citenamefont {Dai}, \citenamefont {Cai},
		\citenamefont {Shi}, \citenamefont {Zhu}, \citenamefont {Guo}, \citenamefont
		{Cao}, \citenamefont {Liu}, \citenamefont {Fert}, \citenamefont {Kioussis},
		\citenamefont {Wang}, \citenamefont {Zhang.},\ and\ \citenamefont
		{Zhao}}]{arXiv02458}%
	\BibitemOpen
	\bibfield  {author} {\bibinfo {author} {\bibfnamefont {S.}~\bibnamefont
			{Xu}}, \bibinfo {author} {\bibfnamefont {Z.}~\bibnamefont {Zhang}}, \bibinfo
		{author} {\bibfnamefont {F.}~\bibnamefont {Mahfouzi}}, \bibinfo {author}
		{\bibfnamefont {Y.}~\bibnamefont {Huang}}, \bibinfo {author} {\bibfnamefont
			{H.}~\bibnamefont {Cheng}}, \bibinfo {author} {\bibfnamefont
			{B.}~\bibnamefont {Dai}}, \bibinfo {author} {\bibfnamefont {W.}~\bibnamefont
			{Cai}}, \bibinfo {author} {\bibfnamefont {K.}~\bibnamefont {Shi}}, \bibinfo
		{author} {\bibfnamefont {D.}~\bibnamefont {Zhu}}, \bibinfo {author}
		{\bibfnamefont {Z.}~\bibnamefont {Guo}}, \bibinfo {author} {\bibfnamefont
			{C.}~\bibnamefont {Cao}}, \bibinfo {author} {\bibfnamefont {Y.}~\bibnamefont
			{Liu}}, \bibinfo {author} {\bibfnamefont {A.}~\bibnamefont {Fert}}, \bibinfo
		{author} {\bibfnamefont {N.}~\bibnamefont {Kioussis}}, \bibinfo {author}
		{\bibfnamefont {K.~L.}\ \bibnamefont {Wang}}, \bibinfo {author}
		{\bibfnamefont {Y.}~\bibnamefont {Zhang.}},\ and\ \bibinfo {author}
		{\bibfnamefont {W.}~\bibnamefont {Zhao}},\ }\href@noop {} {\bibinfo {title}
		{Spin-flop magnetoresistance in a collinear antiferromagnetic tunnel
			junction}} (\bibinfo {year} {2023}),\ \Eprint
	{https://arxiv.org/abs/2311.02458} {arXiv:2311.02458} \BibitemShut {NoStop}%
	\bibitem [{\citenamefont {Gurung}\ \emph {et~al.}(2023)\citenamefont {Gurung},
		\citenamefont {Shao},\ and\ \citenamefont {Tsymbal}}]{arXiv03026}%
	\BibitemOpen
	\bibfield  {author} {\bibinfo {author} {\bibfnamefont {G.}~\bibnamefont
			{Gurung}}, \bibinfo {author} {\bibfnamefont {D.-F.}\ \bibnamefont {Shao}},\
		and\ \bibinfo {author} {\bibfnamefont {E.~Y.}\ \bibnamefont {Tsymbal}},\
	}\href@noop {} {\bibinfo {title} {Extraordinary tunneling magnetoresistance
			in antiferromagnetic tunnel junctions with antiperovskite electrodes}}
	(\bibinfo {year} {2023}),\ \Eprint {https://arxiv.org/abs/2306.03026}
	{arXiv:2306.03026} \BibitemShut {NoStop}%
	\bibitem [{\citenamefont {Yuan}\ \emph {et~al.}(2020)\citenamefont {Yuan},
		\citenamefont {Wang}, \citenamefont {Luo}, \citenamefont {Rashba},\ and\
		\citenamefont {Zunger}}]{PRB-102-2020}%
	\BibitemOpen
	\bibfield  {author} {\bibinfo {author} {\bibfnamefont {L.-D.}\ \bibnamefont
			{Yuan}}, \bibinfo {author} {\bibfnamefont {Z.}~\bibnamefont {Wang}}, \bibinfo
		{author} {\bibfnamefont {J.-W.}\ \bibnamefont {Luo}}, \bibinfo {author}
		{\bibfnamefont {E.~I.}\ \bibnamefont {Rashba}},\ and\ \bibinfo {author}
		{\bibfnamefont {A.}~\bibnamefont {Zunger}},\ }\bibfield  {title} {\bibinfo
		{title} {Giant momentum-dependent spin splitting in centrosymmetric low-{$Z$}
			antiferromagnets},\ }\href {https://doi.org/10.1103/PhysRevB.102.014422}
	{\bibfield  {journal} {\bibinfo  {journal} {Phys. Rev. B}\ }\textbf {\bibinfo
			{volume} {102}},\ \bibinfo {pages} {014422} (\bibinfo {year}
		{2020})}\BibitemShut {NoStop}%
	\bibitem [{\citenamefont {Yuan}\ \emph {et~al.}(2021)\citenamefont {Yuan},
		\citenamefont {Wang}, \citenamefont {Luo},\ and\ \citenamefont
		{Zunger}}]{PRM-2021}%
	\BibitemOpen
	\bibfield  {author} {\bibinfo {author} {\bibfnamefont {L.-D.}\ \bibnamefont
			{Yuan}}, \bibinfo {author} {\bibfnamefont {Z.}~\bibnamefont {Wang}}, \bibinfo
		{author} {\bibfnamefont {J.-W.}\ \bibnamefont {Luo}},\ and\ \bibinfo {author}
		{\bibfnamefont {A.}~\bibnamefont {Zunger}},\ }\bibfield  {title} {\bibinfo
		{title} {Prediction of low-{Z} collinear and noncollinear antiferromagnetic
			compounds having momentum-dependent spin splitting even without spin-orbit
			coupling},\ }\href {https://doi.org/10.1103/PhysRevMaterials.5.014409}
	{\bibfield  {journal} {\bibinfo  {journal} {Phys. Rev. Mater.}\ }\textbf
		{\bibinfo {volume} {5}},\ \bibinfo {pages} {014409} (\bibinfo {year}
		{2021})}\BibitemShut {NoStop}%
	\bibitem [{\citenamefont {Hayami}\ \emph {et~al.}(2019)\citenamefont {Hayami},
		\citenamefont {Yanagi},\ and\ \citenamefont {Kusunose}}]{JPSJ-2019}%
	\BibitemOpen
	\bibfield  {author} {\bibinfo {author} {\bibfnamefont {S.}~\bibnamefont
			{Hayami}}, \bibinfo {author} {\bibfnamefont {Y.}~\bibnamefont {Yanagi}},\
		and\ \bibinfo {author} {\bibfnamefont {H.}~\bibnamefont {Kusunose}},\
	}\bibfield  {title} {\bibinfo {title} {Momentum-dependent spin splitting by
			collinear antiferromagnetic ordering},\ }\href
	{https://doi.org/10.7566/JPSJ.88.123702} {\bibfield  {journal} {\bibinfo
			{journal} {J. Phys. Soc. Jpn.}\ }\textbf {\bibinfo {volume} {88}},\ \bibinfo
		{pages} {123702} (\bibinfo {year} {2019})}\BibitemShut {NoStop}%
	\bibitem [{SM()}]{SM}%
	\BibitemOpen
	\href@noop {} {\bibinfo  {journal} {{See Supplemental Material at [url] for
				details of first-principles calculations; the spin splitting band structures
				of FeF$_2$, the conduction channels of RuO$_2$, IrO$_2$ and CrO$_2$; the
				atomic structures and transmission coefficient distributions of
				RuO$_2$(001)/FeF$_2$/IrO$_2$ MTJ; the atomic and electronic structures, and
				the lowest decay rate distributions of MnF$_2$; simple model of the
				dependence of TMR ratio on the spin filtering barrier thickness}, which
			includes Refs.
			\cite{PRB-54-1996,PRB-50-1994,PRB-59-1999,PRL-77-1996,PRB-44-1991,PRB-57-1998,AC-1997,PRL-118-2017,PRL-86-2001,PRB-69-2004-1,PRL-88-2002,PRB-63-2001-3}}\
	}\BibitemShut {NoStop}%
	\bibitem [{\citenamefont {Belashchenko}\ \emph {et~al.}(2004)\citenamefont
		{Belashchenko}, \citenamefont {Tsymbal}, \citenamefont {van Schilfgaarde},
		\citenamefont {Stewart}, \citenamefont {Oleynik},\ and\ \citenamefont
		{Jaswal}}]{PRB-69-2004-2}%
	\BibitemOpen
	\bibfield  {journal} {  }\bibfield  {author} {\bibinfo {author} {\bibfnamefont
			{K.~D.}\ \bibnamefont {Belashchenko}}, \bibinfo {author} {\bibfnamefont
			{E.~Y.}\ \bibnamefont {Tsymbal}}, \bibinfo {author} {\bibfnamefont
			{M.}~\bibnamefont {van Schilfgaarde}}, \bibinfo {author} {\bibfnamefont
			{D.~A.}\ \bibnamefont {Stewart}}, \bibinfo {author} {\bibfnamefont {I.~I.}\
			\bibnamefont {Oleynik}},\ and\ \bibinfo {author} {\bibfnamefont {S.~S.}\
			\bibnamefont {Jaswal}},\ }\bibfield  {title} {\bibinfo {title} {Effect of
			interface bonding on spin-dependent tunneling from the oxidized {Co}
			surface},\ }\href {https://doi.org/10.1103/PhysRevB.69.174408} {\bibfield
		{journal} {\bibinfo  {journal} {Phys. Rev. B}\ }\textbf {\bibinfo {volume}
			{69}},\ \bibinfo {pages} {174408} (\bibinfo {year} {2004})}\BibitemShut
	{NoStop}%
	\bibitem [{\citenamefont {Velev}\ \emph {et~al.}(2007)\citenamefont {Velev},
		\citenamefont {Duan}, \citenamefont {Belashchenko}, \citenamefont {Jaswal},\
		and\ \citenamefont {Tsymbal}}]{PRL-98-2007}%
	\BibitemOpen
	\bibfield  {author} {\bibinfo {author} {\bibfnamefont {J.~P.}\ \bibnamefont
			{Velev}}, \bibinfo {author} {\bibfnamefont {C.-G.}\ \bibnamefont {Duan}},
		\bibinfo {author} {\bibfnamefont {K.~D.}\ \bibnamefont {Belashchenko}},
		\bibinfo {author} {\bibfnamefont {S.~S.}\ \bibnamefont {Jaswal}},\ and\
		\bibinfo {author} {\bibfnamefont {E.~Y.}\ \bibnamefont {Tsymbal}},\
	}\bibfield  {title} {\bibinfo {title} {Effect of ferroelectricity on electron
			transport in {$\mathrm{Pt}/{\mathrm{BaTiO}}_{3}/\mathrm{Pt}$} tunnel
			junctions},\ }\href {https://doi.org/10.1103/PhysRevLett.98.137201}
	{\bibfield  {journal} {\bibinfo  {journal} {Phys. Rev. Lett.}\ }\textbf
		{\bibinfo {volume} {98}},\ \bibinfo {pages} {137201} (\bibinfo {year}
		{2007})}\BibitemShut {NoStop}%
	\bibitem [{\citenamefont {Chi}\ \emph {et~al.}(2024)\citenamefont {Chi},
		\citenamefont {Jiang}, \citenamefont {Zhu}, \citenamefont {Yu}, \citenamefont
		{Wan}, \citenamefont {Zhang},\ and\ \citenamefont {Han}}]{PRA-21-2024}%
	\BibitemOpen
	\bibfield  {author} {\bibinfo {author} {\bibfnamefont {B.}~\bibnamefont
			{Chi}}, \bibinfo {author} {\bibfnamefont {L.}~\bibnamefont {Jiang}}, \bibinfo
		{author} {\bibfnamefont {Y.}~\bibnamefont {Zhu}}, \bibinfo {author}
		{\bibfnamefont {G.}~\bibnamefont {Yu}}, \bibinfo {author} {\bibfnamefont
			{C.}~\bibnamefont {Wan}}, \bibinfo {author} {\bibfnamefont {J.}~\bibnamefont
			{Zhang}},\ and\ \bibinfo {author} {\bibfnamefont {X.}~\bibnamefont {Han}},\
	}\bibfield  {title} {\bibinfo {title} {Crystal-facet-oriented altermagnets
			for detecting ferromagnetic and antiferromagnetic states by giant tunneling
			magnetoresistance},\ }\href
	{https://doi.org/10.1103/PhysRevApplied.21.034038} {\bibfield  {journal}
		{\bibinfo  {journal} {Phys. Rev. Appl.}\ }\textbf {\bibinfo {volume} {21}},\
		\bibinfo {pages} {034038} (\bibinfo {year} {2024})}\BibitemShut {NoStop}%
	\bibitem [{\citenamefont {Samanta}\ \emph {et~al.}(2024)\citenamefont
		{Samanta}, \citenamefont {Jiang}, \citenamefont {Paudel}, \citenamefont
		{Shao},\ and\ \citenamefont {Tsymbal}}]{PRB-109-2024}%
	\BibitemOpen
	\bibfield  {author} {\bibinfo {author} {\bibfnamefont {K.}~\bibnamefont
			{Samanta}}, \bibinfo {author} {\bibfnamefont {Y.-Y.}\ \bibnamefont {Jiang}},
		\bibinfo {author} {\bibfnamefont {T.~R.}\ \bibnamefont {Paudel}}, \bibinfo
		{author} {\bibfnamefont {D.-F.}\ \bibnamefont {Shao}},\ and\ \bibinfo
		{author} {\bibfnamefont {E.~Y.}\ \bibnamefont {Tsymbal}},\ }\bibfield
	{title} {\bibinfo {title} {Tunneling magnetoresistance in magnetic tunnel
			junctions with a single ferromagnetic electrode},\ }\href
	{https://doi.org/10.1103/PhysRevB.109.174407} {\bibfield  {journal} {\bibinfo
			{journal} {Phys. Rev. B}\ }\textbf {\bibinfo {volume} {109}},\ \bibinfo
		{pages} {174407} (\bibinfo {year} {2024})}\BibitemShut {NoStop}%
	\bibitem [{\citenamefont {Santos-Ortiz}\ \emph {et~al.}(2013)\citenamefont
		{Santos-Ortiz}, \citenamefont {Volkov}, \citenamefont {Schmid}, \citenamefont
		{Kuo}, \citenamefont {Kisslinger}, \citenamefont {Nag}, \citenamefont
		{Banerjee}, \citenamefont {Zhu},\ and\ \citenamefont
		{Shepherd}}]{ACSAMI-2013}%
	\BibitemOpen
	\bibfield  {author} {\bibinfo {author} {\bibfnamefont {R.}~\bibnamefont
			{Santos-Ortiz}}, \bibinfo {author} {\bibfnamefont {V.}~\bibnamefont
			{Volkov}}, \bibinfo {author} {\bibfnamefont {S.}~\bibnamefont {Schmid}},
		\bibinfo {author} {\bibfnamefont {F.-L.}\ \bibnamefont {Kuo}}, \bibinfo
		{author} {\bibfnamefont {K.}~\bibnamefont {Kisslinger}}, \bibinfo {author}
		{\bibfnamefont {S.}~\bibnamefont {Nag}}, \bibinfo {author} {\bibfnamefont
			{R.}~\bibnamefont {Banerjee}}, \bibinfo {author} {\bibfnamefont
			{Y.}~\bibnamefont {Zhu}},\ and\ \bibinfo {author} {\bibfnamefont {N.~D.}\
			\bibnamefont {Shepherd}},\ }\bibfield  {title} {\bibinfo {title}
		{Microstructure and electronic band structure of pulsed laser deposited iron
			fluoride thin film for battery electrodes},\ }\href
	{https://doi.org/10.1021/am3017569} {\bibfield  {journal} {\bibinfo
			{journal} {ACS Appl. Mater. Interfaces}\ }\textbf {\bibinfo {volume} {5}},\
		\bibinfo {pages} {2387} (\bibinfo {year} {2013})}\BibitemShut {NoStop}%
	\bibitem [{\citenamefont {Song}\ \emph {et~al.}(2018)\citenamefont {Song},
		\citenamefont {Cai}, \citenamefont {Tu}, \citenamefont {Zhang}, \citenamefont
		{Huang}, \citenamefont {Wilson}, \citenamefont {Seyler}, \citenamefont {Zhu},
		\citenamefont {Taniguchi}, \citenamefont {Watanabe}, \citenamefont {McGuire},
		\citenamefont {Cobden}, \citenamefont {Xiao}, \citenamefont {Yao},\ and\
		\citenamefont {Xu}}]{S-2018}%
	\BibitemOpen
	\bibfield  {author} {\bibinfo {author} {\bibfnamefont {T.}~\bibnamefont
			{Song}}, \bibinfo {author} {\bibfnamefont {X.}~\bibnamefont {Cai}}, \bibinfo
		{author} {\bibfnamefont {M.~W.-Y.}\ \bibnamefont {Tu}}, \bibinfo {author}
		{\bibfnamefont {X.}~\bibnamefont {Zhang}}, \bibinfo {author} {\bibfnamefont
			{B.}~\bibnamefont {Huang}}, \bibinfo {author} {\bibfnamefont {N.~P.}\
			\bibnamefont {Wilson}}, \bibinfo {author} {\bibfnamefont {K.~L.}\
			\bibnamefont {Seyler}}, \bibinfo {author} {\bibfnamefont {L.}~\bibnamefont
			{Zhu}}, \bibinfo {author} {\bibfnamefont {T.}~\bibnamefont {Taniguchi}},
		\bibinfo {author} {\bibfnamefont {K.}~\bibnamefont {Watanabe}}, \bibinfo
		{author} {\bibfnamefont {M.~A.}\ \bibnamefont {McGuire}}, \bibinfo {author}
		{\bibfnamefont {D.~H.}\ \bibnamefont {Cobden}}, \bibinfo {author}
		{\bibfnamefont {D.}~\bibnamefont {Xiao}}, \bibinfo {author} {\bibfnamefont
			{W.}~\bibnamefont {Yao}},\ and\ \bibinfo {author} {\bibfnamefont
			{X.}~\bibnamefont {Xu}},\ }\bibfield  {title} {\bibinfo {title} {Giant
			tunneling magnetoresistance in spin-filter van der waals heterostructures},\
	}\href {https://doi.org/10.1126/science.aar4851} {\bibfield  {journal}
		{\bibinfo  {journal} {Science}\ }\textbf {\bibinfo {volume} {360}},\ \bibinfo
		{pages} {1214} (\bibinfo {year} {2018})}\BibitemShut {NoStop}%
	\bibitem [{\citenamefont {Song}\ \emph {et~al.}(2019)\citenamefont {Song},
		\citenamefont {Tu}, \citenamefont {Carnahan}, \citenamefont {Cai},
		\citenamefont {Taniguchi}, \citenamefont {Watanabe}, \citenamefont {McGuire},
		\citenamefont {Cobden}, \citenamefont {Xiao}, \citenamefont {Yao},\ and\
		\citenamefont {Xu}}]{NL-2019}%
	\BibitemOpen
	\bibfield  {author} {\bibinfo {author} {\bibfnamefont {T.}~\bibnamefont
			{Song}}, \bibinfo {author} {\bibfnamefont {M.~W.-Y.}\ \bibnamefont {Tu}},
		\bibinfo {author} {\bibfnamefont {C.}~\bibnamefont {Carnahan}}, \bibinfo
		{author} {\bibfnamefont {X.}~\bibnamefont {Cai}}, \bibinfo {author}
		{\bibfnamefont {T.}~\bibnamefont {Taniguchi}}, \bibinfo {author}
		{\bibfnamefont {K.}~\bibnamefont {Watanabe}}, \bibinfo {author}
		{\bibfnamefont {M.~A.}\ \bibnamefont {McGuire}}, \bibinfo {author}
		{\bibfnamefont {D.~H.}\ \bibnamefont {Cobden}}, \bibinfo {author}
		{\bibfnamefont {D.}~\bibnamefont {Xiao}}, \bibinfo {author} {\bibfnamefont
			{W.}~\bibnamefont {Yao}},\ and\ \bibinfo {author} {\bibfnamefont
			{X.}~\bibnamefont {Xu}},\ }\bibfield  {title} {\bibinfo {title} {Voltage
			control of a van der waals spin-filter magnetic tunnel junction},\ }\href
	{https://doi.org/10.1021/acs.nanolett.8b04160} {\bibfield  {journal}
		{\bibinfo  {journal} {Nano Lett.}\ }\textbf {\bibinfo {volume} {19}},\
		\bibinfo {pages} {915} (\bibinfo {year} {2019})}\BibitemShut {NoStop}%
	\bibitem [{\citenamefont {Momma}\ and\ \citenamefont {Izumi}(2011)}]{JAC-2011}%
	\BibitemOpen
	\bibfield  {author} {\bibinfo {author} {\bibfnamefont {K.}~\bibnamefont
			{Momma}}\ and\ \bibinfo {author} {\bibfnamefont {F.}~\bibnamefont {Izumi}},\
	}\bibfield  {title} {\bibinfo {title} {{{\it VESTA3} for three-dimensional
				visualization of crystal, volumetric and morphology data}},\ }\href
	{https://doi.org/10.1107/S0021889811038970} {\bibfield  {journal} {\bibinfo
			{journal} {J. Appl. Cryst.}\ }\textbf {\bibinfo {volume} {44}},\ \bibinfo
		{pages} {1272} (\bibinfo {year} {2011})}\BibitemShut {NoStop}%
	\bibitem [{\citenamefont {Kresse}\ and\ \citenamefont
		{Furthm\"uller}(1996)}]{PRB-54-1996}%
	\BibitemOpen
	\bibfield  {author} {\bibinfo {author} {\bibfnamefont {G.}~\bibnamefont
			{Kresse}}\ and\ \bibinfo {author} {\bibfnamefont {J.}~\bibnamefont
			{Furthm\"uller}},\ }\bibfield  {title} {\bibinfo {title} {Efficient iterative
			schemes for ab initio total-energy calculations using a plane-wave basis
			set},\ }\href {https://doi.org/10.1103/PhysRevB.54.11169} {\bibfield
		{journal} {\bibinfo  {journal} {Phys. Rev. B}\ }\textbf {\bibinfo {volume}
			{54}},\ \bibinfo {pages} {11169} (\bibinfo {year} {1996})}\BibitemShut
	{NoStop}%
	\bibitem [{\citenamefont {Bl\"ochl}(1994)}]{PRB-50-1994}%
	\BibitemOpen
	\bibfield  {author} {\bibinfo {author} {\bibfnamefont {P.~E.}\ \bibnamefont
			{Bl\"ochl}},\ }\bibfield  {title} {\bibinfo {title} {Projector augmented-wave
			method},\ }\href {https://doi.org/10.1103/PhysRevB.50.17953} {\bibfield
		{journal} {\bibinfo  {journal} {Phys. Rev. B}\ }\textbf {\bibinfo {volume}
			{50}},\ \bibinfo {pages} {17953} (\bibinfo {year} {1994})}\BibitemShut
	{NoStop}%
	\bibitem [{\citenamefont {Kresse}\ and\ \citenamefont
		{Joubert}(1999)}]{PRB-59-1999}%
	\BibitemOpen
	\bibfield  {author} {\bibinfo {author} {\bibfnamefont {G.}~\bibnamefont
			{Kresse}}\ and\ \bibinfo {author} {\bibfnamefont {D.}~\bibnamefont
			{Joubert}},\ }\bibfield  {title} {\bibinfo {title} {From ultrasoft
			pseudopotentials to the projector augmented-wave method},\ }\href
	{https://doi.org/10.1103/PhysRevB.59.1758} {\bibfield  {journal} {\bibinfo
			{journal} {Phys. Rev. B}\ }\textbf {\bibinfo {volume} {59}},\ \bibinfo
		{pages} {1758} (\bibinfo {year} {1999})}\BibitemShut {NoStop}%
	\bibitem [{\citenamefont {Perdew}\ \emph {et~al.}(1996)\citenamefont {Perdew},
		\citenamefont {Burke},\ and\ \citenamefont {Ernzerhof}}]{PRL-77-1996}%
	\BibitemOpen
	\bibfield  {author} {\bibinfo {author} {\bibfnamefont {J.~P.}\ \bibnamefont
			{Perdew}}, \bibinfo {author} {\bibfnamefont {K.}~\bibnamefont {Burke}},\ and\
		\bibinfo {author} {\bibfnamefont {M.}~\bibnamefont {Ernzerhof}},\ }\bibfield
	{title} {\bibinfo {title} {Generalized gradient approximation made simple},\
	}\href {https://doi.org/10.1103/PhysRevLett.77.3865} {\bibfield  {journal}
		{\bibinfo  {journal} {Phys. Rev. Lett.}\ }\textbf {\bibinfo {volume} {77}},\
		\bibinfo {pages} {3865} (\bibinfo {year} {1996})}\BibitemShut {NoStop}%
	\bibitem [{\citenamefont {Anisimov}\ \emph {et~al.}(1991)\citenamefont
		{Anisimov}, \citenamefont {Zaanen},\ and\ \citenamefont
		{Andersen}}]{PRB-44-1991}%
	\BibitemOpen
	\bibfield  {author} {\bibinfo {author} {\bibfnamefont {V.~I.}\ \bibnamefont
			{Anisimov}}, \bibinfo {author} {\bibfnamefont {J.}~\bibnamefont {Zaanen}},\
		and\ \bibinfo {author} {\bibfnamefont {O.~K.}\ \bibnamefont {Andersen}},\
	}\bibfield  {title} {\bibinfo {title} {Band theory and {Mott} insulators:
			Hubbard {$U$} instead of {Stoner} {I}},\ }\href
	{https://doi.org/10.1103/PhysRevB.44.943} {\bibfield  {journal} {\bibinfo
			{journal} {Phys. Rev. B}\ }\textbf {\bibinfo {volume} {44}},\ \bibinfo
		{pages} {943} (\bibinfo {year} {1991})}\BibitemShut {NoStop}%
	\bibitem [{\citenamefont {Dudarev}\ \emph {et~al.}(1998)\citenamefont
		{Dudarev}, \citenamefont {Botton}, \citenamefont {Savrasov}, \citenamefont
		{Humphreys},\ and\ \citenamefont {Sutton}}]{PRB-57-1998}%
	\BibitemOpen
	\bibfield  {author} {\bibinfo {author} {\bibfnamefont {S.~L.}\ \bibnamefont
			{Dudarev}}, \bibinfo {author} {\bibfnamefont {G.~A.}\ \bibnamefont {Botton}},
		\bibinfo {author} {\bibfnamefont {S.~Y.}\ \bibnamefont {Savrasov}}, \bibinfo
		{author} {\bibfnamefont {C.~J.}\ \bibnamefont {Humphreys}},\ and\ \bibinfo
		{author} {\bibfnamefont {A.~P.}\ \bibnamefont {Sutton}},\ }\bibfield  {title}
	{\bibinfo {title} {Electron-energy-loss spectra and the structural stability
			of nickel oxide: An {LSDA+U} study},\ }\href
	{https://doi.org/10.1103/PhysRevB.57.1505} {\bibfield  {journal} {\bibinfo
			{journal} {Phys. Rev. B}\ }\textbf {\bibinfo {volume} {57}},\ \bibinfo
		{pages} {1505} (\bibinfo {year} {1998})}\BibitemShut {NoStop}%
	\bibitem [{\citenamefont {Haines}\ \emph {et~al.}(1997)\citenamefont {Haines},
		\citenamefont {L{\'{e}}ger}, \citenamefont {Schulte},\ and\ \citenamefont
		{Hull}}]{AC-1997}%
	\BibitemOpen
	\bibfield  {author} {\bibinfo {author} {\bibfnamefont {J.}~\bibnamefont
			{Haines}}, \bibinfo {author} {\bibfnamefont {J.~M.}\ \bibnamefont
			{L{\'{e}}ger}}, \bibinfo {author} {\bibfnamefont {O.}~\bibnamefont
			{Schulte}},\ and\ \bibinfo {author} {\bibfnamefont {S.}~\bibnamefont
			{Hull}},\ }\bibfield  {title} {\bibinfo {title} {{Neutron Diffraction Study
				of the Ambient-Pressure, Rutile-Type and the High-Pressure, CaCl${\sb
					2}$-Type Phases of Ruthenium Dioxide}},\ }\href
	{https://doi.org/10.1107/S0108768197008094} {\bibfield  {journal} {\bibinfo
			{journal} {Acta Crystallogr. Sect. B: Struct. Sci.}\ }\textbf {\bibinfo
			{volume} {53}},\ \bibinfo {pages} {880} (\bibinfo {year} {1997})}\BibitemShut
	{NoStop}%
	\bibitem [{\citenamefont {Berlijn}\ \emph {et~al.}(2017)\citenamefont
		{Berlijn}, \citenamefont {Snijders}, \citenamefont {Delaire}, \citenamefont
		{Zhou}, \citenamefont {Maier}, \citenamefont {Cao}, \citenamefont {Chi},
		\citenamefont {Matsuda}, \citenamefont {Wang}, \citenamefont {Koehler},
		\citenamefont {Kent},\ and\ \citenamefont {Weitering}}]{PRL-118-2017}%
	\BibitemOpen
	\bibfield  {author} {\bibinfo {author} {\bibfnamefont {T.}~\bibnamefont
			{Berlijn}}, \bibinfo {author} {\bibfnamefont {P.~C.}\ \bibnamefont
			{Snijders}}, \bibinfo {author} {\bibfnamefont {O.}~\bibnamefont {Delaire}},
		\bibinfo {author} {\bibfnamefont {H.-D.}\ \bibnamefont {Zhou}}, \bibinfo
		{author} {\bibfnamefont {T.~A.}\ \bibnamefont {Maier}}, \bibinfo {author}
		{\bibfnamefont {H.-B.}\ \bibnamefont {Cao}}, \bibinfo {author} {\bibfnamefont
			{S.-X.}\ \bibnamefont {Chi}}, \bibinfo {author} {\bibfnamefont
			{M.}~\bibnamefont {Matsuda}}, \bibinfo {author} {\bibfnamefont
			{Y.}~\bibnamefont {Wang}}, \bibinfo {author} {\bibfnamefont {M.~R.}\
			\bibnamefont {Koehler}}, \bibinfo {author} {\bibfnamefont {P.~R.~C.}\
			\bibnamefont {Kent}},\ and\ \bibinfo {author} {\bibfnamefont {H.~H.}\
			\bibnamefont {Weitering}},\ }\bibfield  {title} {\bibinfo {title} {Itinerant
			antirromagnetism in {${\mathrm{RuO}}_{2}$}},\ }\href
	{https://doi.org/10.1103/PhysRevLett.118.077201} {\bibfield  {journal}
		{\bibinfo  {journal} {Phys. Rev. Lett.}\ }\textbf {\bibinfo {volume} {118}},\
		\bibinfo {pages} {077201} (\bibinfo {year} {2017})}\BibitemShut {NoStop}%
	\bibitem [{\citenamefont {Strempfer}\ \emph {et~al.}(2001)\citenamefont
		{Strempfer}, \citenamefont {R\"utt},\ and\ \citenamefont
		{Jauch}}]{PRL-86-2001}%
	\BibitemOpen
	\bibfield  {author} {\bibinfo {author} {\bibfnamefont {J.}~\bibnamefont
			{Strempfer}}, \bibinfo {author} {\bibfnamefont {U.}~\bibnamefont {R\"utt}},\
		and\ \bibinfo {author} {\bibfnamefont {W.}~\bibnamefont {Jauch}},\ }\bibfield
	{title} {\bibinfo {title} {Absolute spin magnetic moment of
			{${\mathrm{FeF}}_{2}$} from high energy photon diffraction},\ }\href
	{https://doi.org/10.1103/PhysRevLett.86.3152} {\bibfield  {journal} {\bibinfo
			{journal} {Phys. Rev. Lett.}\ }\textbf {\bibinfo {volume} {86}},\ \bibinfo
		{pages} {3152} (\bibinfo {year} {2001})}\BibitemShut {NoStop}%
	\bibitem [{\citenamefont {Strempfer}\ \emph {et~al.}(2004)\citenamefont
		{Strempfer}, \citenamefont {R\"utt}, \citenamefont {Bayrakci}, \citenamefont
		{Br\"uckel},\ and\ \citenamefont {Jauch}}]{PRB-69-2004-1}%
	\BibitemOpen
	\bibfield  {author} {\bibinfo {author} {\bibfnamefont {J.}~\bibnamefont
			{Strempfer}}, \bibinfo {author} {\bibfnamefont {U.}~\bibnamefont {R\"utt}},
		\bibinfo {author} {\bibfnamefont {S.~P.}\ \bibnamefont {Bayrakci}}, \bibinfo
		{author} {\bibfnamefont {T.}~\bibnamefont {Br\"uckel}},\ and\ \bibinfo
		{author} {\bibfnamefont {W.}~\bibnamefont {Jauch}},\ }\bibfield  {title}
	{\bibinfo {title} {Magnetic properties of transition metal fluorides
			{$M{\mathrm{F}}_{2}$ $(M=\mathrm{Mn},$} {Fe}, {Co}, {Ni}) via high-energy
			photon diffraction},\ }\href {https://doi.org/10.1103/PhysRevB.69.014417}
	{\bibfield  {journal} {\bibinfo  {journal} {Phys. Rev. B}\ }\textbf {\bibinfo
			{volume} {69}},\ \bibinfo {pages} {014417} (\bibinfo {year}
		{2004})}\BibitemShut {NoStop}%
	\bibitem [{\citenamefont {Goering}\ \emph {et~al.}(2002)\citenamefont
		{Goering}, \citenamefont {Bayer}, \citenamefont {Gold}, \citenamefont
		{Sch\"utz}, \citenamefont {Rabe}, \citenamefont {R\"udiger},\ and\
		\citenamefont {G\"untherodt}}]{PRL-88-2002}%
	\BibitemOpen
	\bibfield  {author} {\bibinfo {author} {\bibfnamefont {E.}~\bibnamefont
			{Goering}}, \bibinfo {author} {\bibfnamefont {A.}~\bibnamefont {Bayer}},
		\bibinfo {author} {\bibfnamefont {S.}~\bibnamefont {Gold}}, \bibinfo {author}
		{\bibfnamefont {G.}~\bibnamefont {Sch\"utz}}, \bibinfo {author}
		{\bibfnamefont {M.}~\bibnamefont {Rabe}}, \bibinfo {author} {\bibfnamefont
			{U.}~\bibnamefont {R\"udiger}},\ and\ \bibinfo {author} {\bibfnamefont
			{G.}~\bibnamefont {G\"untherodt}},\ }\bibfield  {title} {\bibinfo {title}
		{{Strong} {Anisotropy} of {Projected} 3$d$ {Moments} in {Epitaxial} {CrO$_2$}
			{Films}},\ }\href {https://doi.org/10.1103/PhysRevLett.88.207203} {\bibfield
		{journal} {\bibinfo  {journal} {Phys. Rev. Lett.}\ }\textbf {\bibinfo
			{volume} {88}},\ \bibinfo {pages} {207203} (\bibinfo {year}
		{2002})}\BibitemShut {NoStop}%
	\bibitem [{\citenamefont {Taylor}\ \emph {et~al.}(2001)\citenamefont {Taylor},
		\citenamefont {Guo},\ and\ \citenamefont {Wang}}]{PRB-63-2001-3}%
	\BibitemOpen
	\bibfield  {author} {\bibinfo {author} {\bibfnamefont {J.}~\bibnamefont
			{Taylor}}, \bibinfo {author} {\bibfnamefont {H.}~\bibnamefont {Guo}},\ and\
		\bibinfo {author} {\bibfnamefont {J.}~\bibnamefont {Wang}},\ }\bibfield
	{title} {\bibinfo {title} {Ab initio modeling of quantum transport properties
			of molecular electronic devices},\ }\href
	{https://doi.org/10.1103/PhysRevB.63.245407} {\bibfield  {journal} {\bibinfo
			{journal} {Phys. Rev. B}\ }\textbf {\bibinfo {volume} {63}},\ \bibinfo
		{pages} {245407} (\bibinfo {year} {2001})}\BibitemShut {NoStop}%
\end{thebibliography}
\end{document}